\documentclass{amsart}

\newcommand {\supplus}{\mathop{{\supset}\llap{\raise 
0.5pt\hbox{\normalfont\small+}\hskip 0.5pt}}} 

\newcommand {\subplus}{\mathop{{\subset}\llap{\raise 
0.5pt\hbox{\normalfont\small+}\hskip 0.5pt}}}  

\newcommand {\Cee}    {{\mathbb  C}}

\newcommand {\Pee}    {{\mathbb  P}}

\newcommand {\Zee}    {{\mathbb  Z}}

\newcommand {\fab}    {{\mathfrak{ab}}} 

\newcommand {\fas}    {{\mathfrak{as}}}
\newcommand {\faut}   {{\mathfrak{aut}}} 
\newcommand {\fb}     {{\mathfrak{b}}}
\newcommand {\fc}    {{\mathfrak{c}}}
\newcommand {\fcg}    {{\mathfrak{cg}}}

\newcommand {\fder}   {{\mathfrak{der}}}   %

\newcommand {\fdg}    {{\mathfrak{dg}}}
\newcommand {\fd}     {{\mathfrak{d}}}

\newcommand {\fg}     {{\mathfrak{g}}}    %
\newcommand {\fgl}    {{\mathfrak{gl}}}  %
\newcommand {\fh}     {{\mathfrak{h}}}
\newcommand {\fhei}   {{\mathfrak{hei}}}
\newcommand {\fk}     {{\mathfrak{k}}}

\newcommand {\fL}     {{\mathfrak{L}}}
\newcommand {\fle}    {{\mathfrak{le}}}
\newcommand {\fm}     {{\mathfrak{m}}}

\newcommand {\fns}     {{\mathfrak{ns}}}
\newcommand {\fo}     {{\mathfrak{o}}}
\newcommand {\fosp}   {{\mathfrak{osp}}}
\newcommand {\fp}    {{\mathfrak{p}}}   %
\newcommand {\fpe}    {{\mathfrak{pe}}}   %

\newcommand {\fpo}    {{\mathfrak{po}}}

\newcommand {\fr}     {{\mathfrak{r}}}

\newcommand {\fs}     {{\mathfrak{s}}}

\newcommand {\fsb}    {{\mathfrak{sb}}}

\newcommand {\fsh}    {{\mathfrak{sh}}}

\newcommand {\fsl}    {{\mathfrak{sl}}}
\newcommand {\fsle}   {{\mathfrak{sle}}}
\newcommand {\fsm}    {{\mathfrak{sm}}}
\newcommand {\fsp}    {{\mathfrak{sp}}}
\newcommand {\fspe}   {{\mathfrak{spe}}}
\newcommand {\fspo}   {{\mathfrak{spo}}}
\newcommand {\fsq}    {{\mathfrak{sq}}}

\newcommand {\fsvect} {{\mathfrak{svect}}}

\newcommand {\fvect}  {{\mathfrak{vect}}}   %
\newcommand {\fvir}   {{\mathfrak{vir}}}

\newcommand {\fwitt}  {{\mathfrak{witt}}}

\newcommand {\fz}     {{\mathfrak{z}}}

\newcommand {\cal} {\mathcal}

\newcommand {\cC}     {{\cal C}}
\newcommand {\cD}     {{\cal D}}

\newcommand {\cF}     {{\cal F}}

\newcommand {\cK}     {{\cal K}}
\newcommand {\cL}     {{\cal L}}
\newcommand {\cM}     {{\cal M}}

\def \opname#1#2%
  {\expandafter\newcommand \csname #1\endcsname {{\mathop{#2}\nolimits}}}

\newcommand{\rmname}[1]
  {\expandafter\newcommand \csname #1\endcsname {{\operatorname{#1}}}}

\newcommand{\rmnameii}[2]
  {\expandafter\newcommand \csname #1\endcsname {{\operatorname{#2}}}}

\rmname{act}
\rmname{Ad}
\rmname{Add}
\rmname{ad}
\rmname{Alt}
\rmname{alt}
\rmname{Ann}
\rmname{antidiag}
\rmname{Ber}
\rmname{ber}
\rmname{Br}
\rmname{card}
\rmname{ch}
\rmname{Char}
\rmname{cem}
\rmname{cj}
\rmname{Cliff}
\rmname{cntr}
\rmname{codim}
\rmname{coind}
\rmname{const}
\rmname{col}
\rmname{cork}
\rmname{cpr}
\rmname{diag}
\rmnameii{Div}{div}
\rmname{Def}
\rmname{Der}
\rmname{Dim}
\rmname{End}
\rmname{Even}
\rmname{Ext}
\rmname{gr}
\rmname{Hom}
\rmname{HT}
\rmnameii{Ht}{ht}
\rmname{hwt}
\rmname{Id}
\rmname{id}
\rmname{ind}
\rmname{Ind}
\rmname{Inf}
\rmname{irr}
\rmname{Le}
\rmname{Lie}
\rmname{lwt}
\rmname{mult}
\rmname{Mor}
\rmname{nm}
\rmname{Ob}
\rmname{Odd}
\rmname{Osc}
\rmname{per}
\rmname{Pic}
\rmname{pr}
\rmname{pro}
\rmname{Prime}
\rmname{Proj}
\rmname{prt}
\rmname{pt}
\rmname{Q}
\rmname{qet}
\rmname{qtr}
\rmname{rd}
\rmname{rk}
\rmname{row}
\rmname{Res}
\rmname{salt}
\rmname{Sch}
\rmname{SBr}
\rmname{scalar}
\rmname{Ser}
\rmname{sign}
\rmname{Smbl}
\rmname{spin}
\rmname{ssym}
\rmname{str}
\rmname{st}
\rmname{sgn}
\rmname{sq}
\rmname{symm}
\rmname{supp}
\rmname{Supp}
\rmname{St}
\rmname{Spec}
\rmname{Spm}
\rmname{tr}
\rmname{vpt}
\rmname{weyl}
\rmname{Weyl}
\rmname{Witt}

\opname{vvol}  {{v\hspace{-0.1ex}o\hspace{-0.02ex}l\/}}
\opname{pnt}  {\text{\normalfont pt}}
\opname{Span} {{Span}}
\opname{slim} {\overline{\lim}}
\opname{Vol}  {{V\hspace{-0.55ex}o\hspace{-0.02ex}l\/}}
\opname{QVol} {{Q\hspace{-0.3ex}V\hspace{-0.55ex}o\hspace{-0.02ex}l\/}}
\opname{PoVol}{{P\hspace{-0.35ex}o\hspace{-0.25ex}V\hspace{-0.55ex}o\hspace{-0.02ex}l\/}}
\opname{BVol} {{B\hspace{-0.2ex}V\hspace{-0.55ex}o\hspace{-0.02ex}l\/}}
\opname{Par}  {{P\hspace{-0.3ex}a\hspace{-0.05ex}r\/}}

\rmname{Mat}
\rmname{Bil}
\rmname{Diff}
\rmname{Ker}
\rmname{Herm}
\rmname{Coker}
\rmname{Conn}
\rmname{Covect}
\rmname{Vect}
\rmname{Int}

\rmnameii {IM} {Im}
\rmnameii {RE} {Re}

\opname{Aut} {{A\hspace{-0.2ex}u\hspace{-0.1ex}t\/}}
\opname{GL} {{G\hspace{-0.3ex}L}}
\opname{SL} {{S\hspace{-0.3ex}L}}
\opname{Exp} {{E\hspace{-0.2ex}x\hspace{-0.1ex}p\/}}
\opname{GQ} {{G\hspace{-0.2ex}Q}}
\opname{OSp} {{O\hspace{-0.25ex}S\hspace{-0.15ex}p\/}}
\opname{Out} {{O\hspace{-0.25ex}u\hspace{-0.15ex}t\/}}
\opname{Spp} {{S\hspace{-0.2ex}p\/}}
\opname{SpO} {{S\hspace{-0.2ex}p\hspace{-0.02ex}O\/}}
\opname{Pe} {{P\hspace{-0.25ex}e\/}}
\opname{SPe} {{S\hspace{-0.25ex}P\hspace{-0.25ex}e\/}}
\opname{Spin} {{S\hspace{-0.25ex}p\hspace{-0.05ex}i\hspace{-0.1ex}n\/}}
\opname{Iso} {{I\hspace{-0.25ex}s\hspace{-0.1ex}o\/}}
\opname{SSPe} {{S\hspace{-0.25ex}S\hspace{-0.15ex}P\hspace{-0.25ex}e\/}}
\opname{PeU} {{P\hspace{-0.25ex}e\hspace{-0.1ex}U\/}}
\opname{QU} {{Q\hspace{-0.15ex}U\/}}
\opname{U} {{U\/}}

\opname{cGQ} {{\cal G \hspace{-0.2em} Q \/}}
\opname{cSL} {{\cal S \hspace{-0.2em} L \/}}
\opname{cGL} {{\cal G \hspace{-0.2em} L \/}}
\opname{cOSp} {{\cal O \hspace{-0.2em} S \hspace{-0.3em} \it p\/}}
\opname{cPe} {{\cal P \hspace{-1.5pt} \it e\/}}
\opname{cVect} {{\cal V \hspace{-1.5pt} \it e\hspace{-0.1ex}c\hspace{-0.1ex}t\/}}
\opname{cVol} {{\cal V \hspace{-1.5pt} \it o\hspace{-0.1ex}l\/}}
\opname{cAut} {{\cal A \hspace{-0.2em} \it u\hspace{-0.1em}t\/}}
\opname{cCovect} {{\cal C \hspace{-1.5pt}
     \it o\hspace{-0.1ex}v\hspace{-0.1ex}e\hspace{-0.1ex}c\hspace{-0.1ex}t\/}}
\opname{CW} {{C\hspace{-0.15ex}W}}

\opname {Ab}   {{\sf Ab}}
\opname {Alg}   {{\sf Alg}}
\opname {ASch}  {{\sf Aff\;Sch}}
\opname {Funct}   {{\sf Funct}}
\opname {Gr}   {{\sf Gr}}
\opname {Grf}  {{{\sf Gr}_f}}
\opname {Mods}   {{\sf mods}}
\opname {Rings}   {{\sf Rings}}
\opname {Salg}   {{\sf Salg}}
\opname {Sets} {{\sf Sets}}
\opname {SSMan} {{\sf SMan}}
\opname {Top}  {{\sf Top}}
\opname {Vebun}   {{\sf Vebun}}

\newcommand {\ev} {{\bar0}}
\newcommand {\od} {{\bar1}}
\newcommand {\eps} {\varepsilon}
\newcommand {\degree}  {{}^\circ}
\newcommand {\tto} {\longrightarrow}
\newcommand {\pder}[1] {{\frac{\partial}{\partial {#1}}}}
\newcommand {\pderf}[2] {{\frac{\partial {#1}}{\partial {#2}}}}

\newcommand {\bcdot}   {\mathbin{\hbox{\raise.4ex\hbox{\bf.}}}} 

\newcommand {\secno} {}

\newtheorem{Theorem}{\secno Theorem}

\newenvironment {th*}[1]
    {\gdef\thname{#1} \begin{thn}}%
    {\end{thn}}
\newtheorem{thn}[Theorem] {\thname}

\theoremstyle{definition}

\newenvironment {ex*}[1]
    {\gdef\thname{#1} \begin{exn}}%
    {\end{exn}}
\newtheorem{exn}[Theorem]{\thname}

\theoremstyle{remark}

\newenvironment {rem*}[1]
    {\gdef\thname{#1} \begin{remn}}%
    {\end{remn}}
\newtheorem{remn}[Theorem]{\thname}

\newcommand {\ssec}{\subsection*}

\begin{document}

\title{Lie superalgebras of string theories}

\author{Pavel Grozman${}^1$, Dimitry Leites${}^1$ and Irina Shchepochkina${}^2$} 

\address{${}^1$Correspondence: D. Leites, Dept. of Math., Univ. of Stockholm,
Roslagsv. 101, Kr\"aftriket hus 6, S-106 91, Stockholm, Sweden; ${}^2$Independent
Univ. of Moscow, Moscow, Russia.}

\keywords {Lie superalgebra, stringy superalgebra, superconformal algebra, Cartan
prolongation, Neveu--Schwarz superalgebra, Ramond superalgebra, Liouville equation.}

\subjclass{17A70, 17B35} 

\begin{abstract} We describe {\it simple} complex Lie superalgbras of
vector fields on \lq\lq supercircles" --- stringy superalgebras. There are four series of
such algebras (one depends on a complex parameter as well as on an integer one) and four
exceptional stringy superalgebras. Two of the exceptional algebras are new in the
literature. 

The 13 of the simple stringy Lie superalgebras are {\it distinguished}: only they have
nontrivial central extensions and since two of the distinguish algebras have 3 nontrivial central
extensions each, there are exactly 16 superizations of
the Liouville action, Schr\"odinger equation, KdV hierarchy, etc. We also present
the three nontrivial cocycles on the $N=4$ extended Neveu--Schwarz and Ramond superalgebras
in terms of  primary fields. 

We also describe the {\it classical} stringy superalgebras, close to the simple ones. One of these
stringy superalgebras is a Kac--Moody superalgebra $\fg(A)$ with a nonsymmetrizable Cartan matrix
$A$. Unlike the Kac--Moody superalgebras of polynomial growth with symmetrizable Cartan matrix, it can
not be interpreted as a central extension of a twisted loop algebra.

In the litterature the stringy superalgebras are often referred to as {\it superconformal} ones. We
discuss how superconformal stringy superalgebras really are.
\end{abstract} 

\thanks{We are thankful for financial support to: the Swedish Institute, NFR and RFBR grant 
95-01187 01, respectively. D.L. is thankful to J.~W.~van de Leur for stimulating
prompts, see sec. 1.9.}

\maketitle

\section*{Introduction}

\ssec{I.1. The discovery of stringy Lie superalgebras} The discovery of
stringy superalgebras was not a one-time job and their further study was not smooth either.
Even the original name of these Lie superalgebras is unfortunate. (Physicists dubbed them
\lq\lq superconformal" in analogy with the stringy Lie algebra $\fwitt$ of conformal
transformations, but, as we will show, not all stringy {\it super}
algebras are superconformal. Besides, the diversity of them requires often to refer to them
using their \lq\lq given names" that describe them more precisely, rather than the common
term. If the common name is needed, nevertheless, then {\it stringy} is, at least, not
selfcontradictory and suggestive.) We give here an intrinsic description borrowed from [Ma].

The physicists who discovered stringy superalgebras ([NS], [R], [Ad]) were primarily interested in
unitary representations, so they started with real algebras which are more difficult to classify than
complex algebras. So they gave a number of examples, not a classification.

Observe also that physicists who studied superstrings were mainly interested in
nontrivial central extensions of \lq\lq superconformal" superalgebras. Only several first
terms of the four series of stringy superalgebras --- the 13 distinguished
superalgebras --- have such extensions, the other algebras were snubbed at. For a review how 
distinguished stringy superalgebras are used in in string theory see [GSW]. For some
other applications see [LX] and [LSX] (where some of the
results are quite unexpected). Mathematically, nondistinguished simple stringy superalgebras
are also of interest, see [CLL], [GL2] and [LS].

Ordered historically, the steps of classification are: [NS] and [R] followed by [Ad], where 4 series of
the stringy superalgebras (without a continuous parameter) and most of the central extensions of the
distinguished superalgebras were found for one real form of each algebra; [FL], where the
complexifications of the algebras from [Ad] were interpreted geometrically and where
the classifications of simple stringy superalgebras and their central extensions expressed
in terms of {\it superfields} were announced. Regrettably, each classification had a gap. During the
past years these gaps were partly filled in by several authors, in this paper the repair is
completed.

Poletaeva [P] in 1983 and, independently, Schoutens [Sc] in 1986, found 3 nontrivial central extensions
of $\fk^{L0}(1|4)$ and $\fk^M(1|4)$, i.e., of the $4$-extended Neveu-Schwarz and Ramond superalgebras
(the importance of [P], whose results were expressed in unconventional terms, was not recognized
in time and the paper was never properly published).

Schwimmer and Seiberg [SS] found a deformation of the divergence-free series. 

In [KvL] the completeness of the list of examples from [FL] amended with the deformation from [SS] was
conjectured and the statements from [FL] and [Sc] on the nontrivial central extensions reproved; [KvL]
contains the first published proof of the classification of the nontrivial central extensions of the
simple  superalgebras considered. 

Other important steps of classification: [K], [L1] and [Sch], where the vectoral Lie
superalgebras with polynomial coefficients are considered, and [Ma], where a characterizatin
of stringy algebras is introduced.

Observe, that after [Sc] there appeared several papers in which only two of the three
central extensions of $\fk^{L0}(1|4)$ and $\fk^M(1|4)$ were recognized; the controversy is
occasioned, presumably, by the insufficiently lucid description of the superalgebras involved and
ensuing confusion between the exceptional simple superlagebra $\fk^{L0}(1|4)$ and
$\fk^{L}(1|4)$, see below. Besides, the cocycles that the physicists need should be expressed in
terms of the primary fields; so far, this was not done.

\ssec{I.2. Our results} Here we define classical {\it stringy} Lie superalgebras, formely
superconformal ones, and announce the conjectural list of all {\it simple} stringy
superalgebras.  We thus repair the classificational result of [FL] with the help of later
discovery [Sch]. Two series of the {\it simple} stringy superalgebras and several exceptional ones, as
well as two nontrivial cocycles, seem to be new. Proof of the completeness of the classification will
be given elsewhere.

We also answer a question of S.~Krivonos: we replace the three cocycles
found in [P] and [Sh] with the cohomologic ones but expressed in terms of primary fields.

\ssec{I.3. Related latest results} This paper is accompanied with several related papers.

1) Real forms. A 1986 result of Serganova [S], completed with the real forms of the
distinguished and simple stringy superalgebras unknown to her at that time. Crucial there is
the discovery of 3, not 2, types of real forms of stringy and Kac--Moody superalgebras, cf.
[S] with [K1], where only two types of real forms of Kac--Moody algebras are recognized.
Another important result of Serganova pertaining here: the discovery of three basic types of
unitarity, one of them with an odd form.

2) Semi-infinite cohomology and the critical dimension. See [LSX].

3) Integrable systems. See [LX].

\ssec{Remark} The results of this paper were obtained in Stockholm in June 1996 (except
4.3, explicit formula 1.1.8 and the answer to Krivonos' question) and delivered at the seminar of
E.~Ivanov, JINR, Dubna (July, 1996) and Voronezh winter school Jan. 12--18, 1997. Kac's questions
concerning exceptional Lie superalgebras in his numerous letters to
I.~Shchepochkina in October-November 1996 encouraged us to struggle with the crashing computer
systems and \TeX. While the \TeX-file was beeing processed, we got a recent preprint [CK] by Cheng
Shun-Jen and V.~Kac, where our example $\fk\fas^L$ is described in different terms. We thank Kac for
the kind letter (sent to Leites) that aknowledges his receiving of a preprint of [Sch] and interesting
preprints [CK] and [K2] (where several results from [L2] get a new perspective).

\section*{\S 0. Background}

\ssec{0.1. Linear algebra in superspaces. Generalities}
Superization has certain subtleties, often disregarded or expressed as in [L],
[L3] or [M]: too briefly. We will dwell on them a bit.

A {\it superspace} is a $\Zee /2$-graded space; for a superspace
$V=V_{\ev}\oplus V_{\od }$ denote by $\Pi (V)$ another copy of the same
superspace: with the shifted parity, i.e., $(\Pi(V))_{\bar i}= V_{\bar i+\od
}$. The {\it superdimension} of $V$ is $\dim V=p+q\varepsilon$, where
$\varepsilon^2=1$ and $p=\dim V_{\ev}$,
$q=\dim V_{\od }$. (Usually, $\dim V$ is expressed as a pair $(p,q)$ or
$p|q$; this obscures the fact that $\dim V\otimes W=\dim V\cdot \dim W$, clear with the help
of $\varepsilon$.)

A superspace structure in $V$ induces the superspace structure in the space $\End
(V)$. A {\it basis of a superspace} is always a basis consisting of {\it
homogeneous} vectors; let
$\Par=(p_1, \dots, p_{\dim V})$ be an ordered collection of their parities.
We call $\Par$ the {\it format} of (the basis of) $V$. A square {\it supermatrix} of format
(size) $\Par$ is a $\dim V\times \dim V$ matrix whose $i$th row and $i$th column are of
the same parity $p_i$. The matrix unit $E_{ij}$ is supposed to be of parity
$p_i+p_j$ and the bracket of supermatrices (of the same format) is defined via Sign Rule: {\it if something of
parity $p$ moves past something of parity $q$ the sign $(-1)^{pq}$ accrues; the
formulas defined on homogeneous elements are extended to arbitrary ones via
linearity}. For example, setting $[X, Y]=XY-(-1)^{p(X)p(Y)}YX$ we get the notion
of the supercommutator and the ensuing notion of the Lie superalgebra (that
satisfies the superskew-commutativity and super Jacobi identity).

We do not usually use the sign $\wedge$ for differential forms on supermanifolds: in what
follows we assume that the exterior differential is odd and the differential forms
constitute a supercommutative superalgebra; still, we keep using it on manifolds, sometimes,
not to diviate too far from conventional notations. 

Usually, $\Par$ is of the form $(\ev , \dots, \ev , \od , \dots, \od )$. Such a
format is called {\it standard}. In this paper we can do without nonstandard formats. But
they are vital in the study of systems of simple roots that the reader might be
interested in connection with applications to $q$-quantization or integrable systems.

The {\it general linear} Lie superalgebra of all supermatrices of size $\Par$ is
denoted by $\fgl(\Par)$; usually, $\fgl(\ev, \dots, \ev, \od, \dots, \od)$ is
abbreviated to $\fgl(\dim V_{\bar 0}|\dim V_{\bar 1})$. Any matrix from
$\fgl(\Par)$ can be expressed as the sum of its even
and odd parts; in the standard format this is the following block expression: 
$$
\begin{pmatrix}A&B\\ C&D\end{pmatrix}=\begin{pmatrix}A&0\\
0&D\end{pmatrix}+\begin{pmatrix}0&B\\ C&0\end{pmatrix},\quad p\left(\begin{pmatrix}A&0\\
0&D\end{pmatrix}\right)=\ev, \; p\left(\begin{pmatrix}0&B\\
C&0\end{pmatrix}\right)=\od.
$$

The {\it supertrace} is the map $\fgl (\Par)\longrightarrow \Cee$,
$(A_{ij})\mapsto \sum (-1)^{p_{i}}A_{ii}$. Since $\str [x, y]=0$, the subsuperspace of
supertraceless matrices constitutes the {\it special linear} Lie subsuperalgebra $\fsl(\Par)$.

{\bf Superalgebras that preserve bilinear forms: two types}. To the linear map $F$ of 
superspaces there corresponds the dual map $F^*$ between the
dual superspaces; if $A$ is the supermatrix corresponding to $F$ in a basis of format
$\Par$, then to $F^*$ the {\it supertransposed} matrix $A^{st}$ corresponds:
$$
(A^{st})_{ij}=(-1)^{(p_{i}+p_{j})(p_{i}+p(A))}A_{ji}.
$$

The supermatrices $X\in\fgl(\Par)$ such that 
$$
X^{st}B+(-1)^{p(X)p(B)}BX=0\quad \text{for a homogeneous matrix $B\in\fgl(\Par)$}
$$
constitute the Lie superalgebra $\faut (B)$ that preserves the bilinear form on $V$ with
matrix $B$. Most popular is the nondegenerate supersymmetric form whose matrix in the
standard format is the canonical form $B_{ev}$ or $B'_{ev}$:
$$
B_{ev}(m|2n)= \begin{pmatrix} 
1_m&0\\
0&J_{2n}
\end{pmatrix},\quad \text{where $J_{2n}=\begin{pmatrix}0&1_n\\-1_n&0\end{pmatrix}$,
or}\; B'_{ev}(m|2n)= \begin{pmatrix} 
\antidiag (1, \dots , 1)&0\\
0&J_{2n}
\end{pmatrix}. 
$$
The usual notation for $\faut (B_{ev}(m|2n))$ is $\fosp(m|2n)$ or $\fosp^{sy}(m|2n)$. Observe
that the passage from $V$ to $\Pi (V)$ sends the supersymmetric forms to
superskew-symmetric ones, preserved by the \lq\lq
symplectico-orthogonal" Lie superalgebra, $\fsp'\fo (2n|m)$ or better $\fosp^{sk}(m|2n)$,
which is isomorphic to
$\fosp^{sy}(m|2n)$ but has a different matrix realization. We never use notation
$\fsp'\fo (2n|m)$ in order not to confuse with the special Poisson
superalgebra.

In the standard format the matrix realizations of these algebras
are: 
$$
\begin{matrix} 
\fosp (m|2n)=\left\{\left (\begin{matrix} E&Y&-X^t\\
X&A&B\\
Y^t&C&-A^t\end{matrix} \right)\right\};\quad \fosp^{sk}(m|2n)=
\left\{\left(\begin{matrix} A&B&X\\
C&-A^t&Y^t\\
Y^t&-X^t&E\end{matrix} \right)\right\}, \\
\text{where}\; 
\left(\begin{matrix} A&B\\
C&-A^t\end{matrix} \right)\in \fsp(2n),\qquad E\in\fo(m)\;
\text{and}\;  {}^t \; \text{is the usual transposition}.\end{matrix} 
$$

A nondegenerate supersymmetric odd bilinear form $B_{odd}(n|n)$ can be
reduced to a canonical form whose matrix in the standard format is 
$J_{2n}$. A canonical form of the superskew odd nondegenerate form in the
standard format is $\Pi_{2n}=\begin{pmatrix} 0&1_n\\1_n&0\end{pmatrix}$.
The usual notation for $\faut (B_{odd}(\Par))$ is $\fpe(\Par)$. The passage
from $V$ to $\Pi (V)$ sends the supersymmetric forms to superskew-symmetric
ones and establishes an isomorphism $\fpe^{sy}(\Par)\cong\fpe^{sk}(\Par)$. This Lie
superalgebra is called, as A.~Weil suggested, {\it periplectic}, i.e., odd-plectic. The
matrix realizations in the standard format of these superalgebras is shorthanded to:
$$
\begin{matrix}
\fpe ^{sy}\ (n)=\left\{\begin{pmatrix} A&B\\
C&-A^t\end{pmatrix}, \; \text{where}\; B=-B^t,\; 
C=C^t\right\};\\
\fpe^{sk}(n)=\left\{\begin{pmatrix}A&B\\ C&-A^t\end{pmatrix}, \;
\text{where}\; B=B^t,\;  C=-C^t\right\}.
\end{matrix}
$$

The {\it special periplectic} superalgebra is $\fspe(n)=\{X\in\fpe(n): \str
X=0\}$.

\ssec{0.2. Vectoral Lie superalgebras. The standard realization} The elements of the
Lie algebra$\cL=\fder\;
\Cee [[u]]$ are considered as vector fields. The Lie algebra $\cL$ has only one maximal
subalgebra
$\cL_0$ of finite codimension (consisting of the fields that vanish at the origin). The
subalgebra $\cL_0$ determines a filtration of $\cL$: set
$$
\cL_{-1}=\cL;\quad \cL_i =\{D\in \cL_{i-1}: [D, \cL]\subset\cL_{i-1}\}\; \text{for
}i\geq 1.
$$
The associated graded Lie algebra $L=\mathop{\oplus}\limits_{i\geq -1}L_i$, where
$L_i=\cL_{i}/\cL_{i+1}$, consists of the vector fields with {\it polynomial} coefficients.

Superization and the passage to a subalgebras of $\fder\; \Cee [[u]]$ brings new phenomena.
Suppose $\cL_0\subset\cL$ is a maximal subalgebra of finite codimension and containing no
ideals of $\cL$. For the Lie superalgebra $\cL=\fder\, \Cee [u, \xi]$ the minimal
$\cL_0$-invariant subspace of
$\cL$ strictly containing $\cL_0$ coincides with $\cL$. Not all the subalgebras $\cL$ of
$\fder\,
\Cee [u, \xi]$ have this property. Let $\cL_{-1}$ be a minimal subspace of
$\cL$ containing $\cL_0$, different from $\cL_0$ and $\cL_0$-invariant. A {\it Weisfeiler
filtration} of $\cL$ is determined by the formula
$$
\cL_{-i-1}=[\cL_{-1},
\cL_{-i}]+\cL_{-i}\;\quad \cL_i =\{D\in \cL_{i-1}: [D, \cL{-1}]\subset\cL_{i-1}\}\; 
\text{for }i>0.
$$
Since the codimension of $\fL_0$ is finite, the filtration takes the form
$$
\cL=\cL_{-d}\supset\dots\cL_{0}\supset\dots \eqno{(0.2)}
$$
for some $d$. This $d$ is the {\it depth} of $\cL$ or of the associated graded Lie
superalgebra
$L$. We call all filtered or graded Lie superalgebras of finite depth {\it vectoral}, i.e.,
realizable with vector fields on a finite dimensional supermanifold. Considering the
subspaces $(0.2)$ as the basis of a topology, we can complete the graded or filtered Lie
superalgebras $L$ or $\cL$; the elements of the completion are the vector fields with formal
power series as coefficients. Though the structure of the graded algebras is easier to
describe, in applications the completed Lie superalgebras are usually needed. 

Unlike Lie algebras, simple vectoral {\it super}algebras possess {\it several} maximal
subalgebras of finite codimension. We describe them, together with the corresponding
gradings, in sec. 0.4.

{\bf 1) General algebras}. Let $x=(u_1, \dots , u_n, \theta_1, \dots ,
\theta_m)$, where the $u_i$ are even indeterminates and the $\theta_j$ are odd ones.
Set $\fvect (n|m)=\fder\; \Cee[x]$; it is called {\it the
general vectoral Lie superalgebra}. \index{$\fvect$ general vectoral Lie
superalgebra}\index{ Lie superalgebra general vectoral}

{\bf 2) Special algebras}. The {\it divergence}\index{divergence} of the field
$D=\sum\limits_if_i\pder{u_{i}} + \sum\limits_j
g_j\pder{\theta_{j}}$ is the function (in our case: a
polynomial, or a series) 
$$
\Div D=\sum\limits_i\pderf{f_{i}}{u_{i}}+
\sum\limits_j (-1)^{p(g_{j})}
\pderf{g_{i}}{\theta_{j}}.
$$

$\bullet$ The Lie superalgebra $\fsvect (n|m)=\{D \in \fvect (n|m): \Div D=0\}$
is called the {\it special} or {\it divergence-free vectoral superalgebra}.
\index{$\fsvect$ general vectoral Lie superalgebra}\index{ Lie superalgebra
special vectoral}\index{ Lie superalgebra
divergence-free}

It is clear that it is also possible to describe $\fsvect(n|m)$ as $\{ D\in \fvect
(n|m): L_D\vvol _x=0\}$, where $\vvol_x$ is the volume form with constant
coefficients in coordinates $x$ and $L_D$ the Lie derivative with respect to
$D$. 

$\bullet$ The Lie superalgebra $\fsvect_{\lambda}(0|m)=\{D \in \fvect (0|m):
\Div (1+\lambda\theta_1\cdot \dots \cdot
\theta_m)D=0\}$ --- the deform of $\fsvect(0|m)$ --- is called the
{\it special} or {\it divergence-free vectoral superalgebra}. Clearly,
$\fsvect_{\lambda}(0|m)\cong \fsvect_{\mu}(0|m)$ for
$\lambda\mu\neq 0$. Observe that $p(\lambda)\equiv m\pmod 2$, i.e., for odd $m$
the parameter of deformation $\lambda$ is odd.

\begin{rem*}{Remark} Sometimes we write $\fvect (x)$ or even $\fvect (V)$ if
$V=\Span(x)$ and use similar notations for the subalgebras of $\fvect$ introduced
below. Algebraists sometimes abbreviate $\fvect (n)$ and $\fs\fvect (n)$ to $W_n$
(in honor of Witt) and $S_n$, respectively.
\end{rem*}

{\bf 3) The algebras that preserve Pfaff
equations and differential 2-forms}. 

$\bullet$ Set $u=(t, p_1, \dots , p_n, q_1, \dots , q_n)$; let
$$
\tilde \alpha_1 = dt +\sum\limits_{1\leq i\leq n}(p_idq_i - q_idp_i)\ +
\sum\limits_{1\leq j\leq m}\theta_jd\theta_j\quad\text{and}\quad 
 \omega_0=d\alpha_1\ .
$$
The form $\tilde \alpha_1$ is called {\it contact}, the form  $\tilde \omega_0$ is called {\it
symplectic}.\index{form differential contact}\index{form differential
symplectic} Sometimes it is more convenient to redenote
the $\theta$'s and set 
$$
\xi_j=\frac{1}{\sqrt{2}}(\theta_{j}-i\theta_{r+j});\quad \eta_j=\frac{1}{
\sqrt{2}}(\theta_{j}+i\theta_{r+j})\; \text{ for}\; j\leq r= [m/2]\; (\text{here}\;
i^2=-1),
\quad
\theta =\theta_{2r+1} 
$$ 
and in place of $\tilde \omega_0$ or $\tilde \alpha_1$ take $\alpha$
and $\omega_0=d\alpha_1$, respectively, where 
$$
\begin{array}{rcl}
\alpha_1=&dt+\sum\limits_{1\leq i\leq n}(p_idq_i-q_idp_i)+
\sum\limits_{1\leq j\leq r}(\xi_jd\eta_j+\eta_jd\xi_j)&
\text{ if }\ m=2r\\
\alpha_1=&dt+\sum\limits_{1\leq i\leq n}(p_idq_i-q_idp_i)+
\sum\limits_{1\leq j\leq r}(\xi_jd\eta_j+\eta_jd\xi_j) +\theta d\theta&\text{ if
}\ m=2r+1.\end{array} 
$$

The Lie superalgebra that preserves the {\it Pfaff equation}
\index{Pfaff equation} $\alpha_1=0$, i.e., the superalgebra
$$
\fk (2n+1|m)=\{ D\in \fvect (2n+1|m): L_D\alpha_1=f_D\alpha_1\}, 
$$
(here $f_D\in \Cee [t, p, q, \xi]$ is a polynomial determined by $D$) is
called the {\it contact superalgebra}.\index{$\fk$ contact superalgebra}
\index{Lie superalgebra contact} The Lie superalgebra
$$
\begin{array}{c}
\fpo (2n|m)=\{ D\in \fk (2n+1|m): L_D\alpha_1=0\}\end{array}
$$
is called the {\it Poisson} superalgebra.\index{$\fpo$ Poisson superalgebra}
(A geometric interpretation of the
Poisson superalgebra: it is the Lie superalgebra that preserves the connection
with form $\alpha$ in the line bundle over a symplectic supermanifold
with the symplectic form $d\alpha$.) 

$\bullet$ Similarly, set $u=q=(q_1, \dots , q_n)$,
let $\theta=(\xi_1, \dots , \xi_n; \tau)$ be odd. Set
$$
\begin{array}{c}
\alpha_0=d\tau+\sum\limits_i(\xi_idq_i+q_id\xi_i), \qquad\qquad
\omega_1=d\alpha_0
\end{array}
$$
and call these forms the {\it odd contact} and {\it periplectic}, 
respectively.\index{form differential contact odd}\index{form differential
periplectic}

The Lie superalgebra that preserves
the Pfaff equation $\alpha_0=0$, i.e., the superalgebra 
$$
\fm (n)=\{ D\in \fvect (n|n+1):L_D\alpha_0=f_D\cdot \alpha_0\}, \; \text{
where }\; f_D\in \Cee [q, \xi, \tau], 
$$
is called the {\it odd contact superalgebra}.\index{$\fm$ contact
superalgebra}

The Lie superalgebra \index{$\fb$ Buttin superalgebra}
$$
\begin{array}{c}
\fb (n)=\{ D\in \fm (n): L_D\alpha_0=0\}\end{array}
$$
is called the {\it Buttin} superalgebra ([L3]). (A geometric interpretation of the
Buttin superalgebra: it is the Lie superalgebra that preserves
the connection with form $\alpha_1$ in the line bundle of rank $\varepsilon$ over a
periplectic supermanifold, i.e., a supermanifold with the periplectic
form $d\alpha_0$.)

The Lie superalgebras
$$
\begin{array}{c}
\fsm (n)=\{ D\in \fm (n): \Div\ D=0\}\ , \ \fsb (n)=\{ D\in
\fb (n):\Div\ D=0\} 
\end{array}
$$
are called the {\it divergence-free} (or {\it special}) {\it odd contact} and
{\it special Buttin} superalgebras, respectively.

\begin{rem*}{Remark} A relation with finite dimensional geometry is as follows.
Clearly, $\ker \alpha_1= \ker \alpha_1$. The restriction of $\omega_0$ to
$\ker 
\alpha_1$ is the orthosymplectic form $B_{ev}(m|2n)$; the restriction of
$\omega_0$ to $\ker \alpha_1$ is $B'_{ev}(m|2n)$. Similarly, the restriction of
$\omega _1$ to $\ker \alpha_0$ is $B_{odd}(n|n)$.
\end{rem*}

\ssec{0.3. Generating functions} A laconic way to describe
$\fk$, $\fm$ and their subalgebras is via generating functions.

$\bullet$ Odd form $\alpha_1$. For $f\in\Cee [t, p, q, \xi]$ set\index{$K_f$
contact vector field} \index{$H_f$ Hamiltonian vector field}: 
$$
K_f=\triangle(f)\pder{t}-H_f +
\pderf{f}{t} E, 
$$
where
$E=\sum\limits_i y_i
\pder{y_{i}}$ (here the $y$ are all the
coordinates except $t$) is the {\it Euler operator} (which counts the degree
with respect to the $y$), $\triangle (f)=2f-E(f)$, and $H_f$ is the
hamiltonian field with Hamiltonian $f$ that preserves $d\alpha_1$: 
$$
H_f=\sum\limits_{i\leq n}(\pderf{f}{p_i}
\pder{q_i}-\pderf{f}{q_i}
\pder{p_i}) -(-1)^{p(f)}\left(\sum\limits_{j\leq m}\pderf{
f}{\theta_j} \pder{\theta_j}\right ) , \; \; f\in \Cee [p,
q, \theta]. 
$$

The choice of the form $\alpha_1$ instead of $\tilde\alpha_1$ only affects the
form of $H_f$ that we give for $m=2k+1$:
$$
H_f=\sum\limits_{i\leq n} (\pderf{f}{p_i}
\pder{q_i}-\pderf{f}{q_i}
\pder{p_i}) -(-1)^{p(f)}\sum\limits_{j\leq
k}(\pderf{f}{\xi_j} \pder{\eta_j}+
\pderf{f}{\eta_j} \pder{\xi_j}+
\pderf{f}{\theta} \pder{\theta}), \;
\; f\in \Cee [p, q, \xi, \eta, \theta]. 
$$

$\bullet$ Even form $\alpha_0$. For $f\in\Cee [q, \xi, \tau]$ set:
$$
M_f=\triangle(f)\pder{\tau}- Le_f
-(-1)^{p(f)} \pderf{f}{\tau} E, 
$$
where $E=\sum\limits_iy_i 
\pder{y_i}$ (here the $y$ are all the coordinates except
$\tau$) is the Euler operator, $\triangle(f)=2f-E(f)$, and
$$
Le_f=\sum\limits_{i\leq n}( \pderf{f}{q_i}\ 
\pder{\xi_i}+(-1)^{p(f)} \pderf{f}{\xi_i}\ 
\pder{q_i}), \; f\in \Cee [q, \xi].
$$
\index{$M_f$ contact vector field} \index{$Le_f$ periplectic vector field} 

Since 
$$
L_{K_f}(\alpha_1)=2 \pderf{f}{t}\alpha_1, \quad\quad
L_{M_f}(\alpha_0)=-(-1)^{p(f)}2 \pderf{
f}{\tau}\alpha_0,\eqno{(0.3)}
$$
it follows that $K_f\in \fk (2n+1|m)$ and $M_f\in \fm (n)$. Observe that
$$
p(Le_f)=p(M_f)=p(f)+\od.
$$

$\bullet$ To the (super)commutators $[K_f, K_g]$ or $[M_f, M_g]$ there
correspond {\it contact brackets}\index{Poisson bracket}\index{contact bracket}
of the generating functions:
$$
[K_f, K_g]=K_{\{f, g\}_{k.b.}};\quad\quad [M_f, M_g]=M_{\{f, g\}_{m.b.}}
$$
The explicit formulas for the contact brackets are as follows. Let us first
define the brackets on functions that do not depend on $t$ (resp. $\tau$).

The {\it Poisson bracket} $\{\cdot , \cdot\}_{P.b.}$ (in the realization with the form
$\omega_0$) is given by the formula 
$$
\{f, g\}_{P.b.}=\sum\limits_{i\leq n}\ \bigg(\pderf{f}{p_i}\ 
\pderf{g}{q_i}-\ \pderf{f}{q_i}\ 
\pderf{g}{p_i}\bigg)-(-1)^{p(f)}\sum\limits_{j\leq m}\ 
\pderf{f}{\theta_j}\ \pderf{g}{\theta_j}
$$
and in the realization with the form
$\omega_0$ for $m=2k+1$ it is given by the formula 
$$
\{f, g\}_{P.b.}=\sum\limits_{i\leq n}\ \bigg(\pderf{f}{p_i}\ 
\pderf{g}{q_i}-\ \pderf{f}{q_i}\ 
\pderf{g}{p_i}\bigg)-(-1)^{p(f)}\bigg[\sum\limits_{j\leq m}( 
\pderf{f}{\xi_j}\ \pderf{
g}{\eta_j}+\pderf{f}{\eta_j}\ \pderf{
g}{\xi_j})+\pderf{f}{\theta}\ \pderf{
g}{\theta}\bigg].
$$

The {\it Buttin bracket} $\{\cdot ,
\cdot\}_{B.b.}$ \index{Buttin bracket $=$ Schouten bracket} is given by the formula
$$
\{ f, g\}_{B.b.}=\sum\limits_{i\leq n}\ \bigg(\pderf{f}{q_i}\ 
\pderf{g}{\xi_i}+(-1)^{p(f)}\ \pderf{f}{\xi_i}\ 
\pderf{g}{q_i}\bigg).
$$

\footnotesize
\begin{rem*}{Remark} The what we call here Buttin bracket was
discovered in pre-super era by Schouten; Buttin proved that this bracket establishes a Lie
superalgebra structure and the interpretation of this superalgebra similar to that of Poisson
algebra was given in [L1]. The {\it Schouten bracket} was originally defined on the superspace
of polyvector fields on a manifold, i.e., on the superspace of sections of the exterior algebra
(over the algebra
$\cF$ of functions) of the tangent bundle,
$\Gamma(\Lambda^{\bcdot}(T(M)))\cong\Lambda^{\bcdot}_\cF(Vect(M))$. The explicit
formula of the Schouten bracket (in which the hatted slot should be ignored, as
usual) is 
$$
[X_1\wedge\dots \wedge\dots \wedge X_k, Y_1\wedge\dots \wedge Y_l]=
\sum_{i, j}(-1)^{i+j}[X_i, Y_j]\wedge X_1\wedge\dots\wedge \hat X_i\wedge 
\dots\wedge X_k\wedge Y_1\wedge\dots\wedge \hat Y_j\wedge \dots \wedge
Y_l.\eqno{(*)}
$$
With the help of Sign Rule we easily superize formula $(*)$ for the case
when manifold $M$ is replaced with supermanifold $\cM$. Let $x$ and $\xi$
be the even and odd coordinates on $\cM$. Setting $\theta_i=\Pi(\partial
{x_{i}})=\check x_{i}$, $q_j=\Pi(\partial {\xi_{j}})=\check \xi_{j}$ we get an
identification of the Schouten bracket of polyvector fields on $\cM$ with the
Buttin bracket of functions on the supermanifold $\check\cM$ whose coordinates are $x,
\xi$ and  $\check x=\Pi(\pder{x})$, $\check \xi= \Pi(\pder{\xi})$; the transformation of $x,
\xi$ induces from that of the checked coordinates. 
\end{rem*}
\normalsize

In terms of the Poisson and Buttin brackets, respectively, the contact
brackets take the form
$$
\{ f, g\}_{k.b.}=\triangle (f)\pderf{g}{t}-\pderf{f}
{t}\triangle (g)-\{ f, g\}_{P.b.}
$$
and, respectively,
$$
\{ f, g\}_{m.b.}=\triangle (f)\pderf{g}{\tau}+(-1)^{p(f)}
\pderf{f}{\tau}\triangle (g)-\{ f, g\}_{B.b.}.
$$

The Lie superalgebras of {\it Hamiltonian fields}\index{Hamiltonian
vector fields} (or {\it Hamiltonian 
superalgebra}) and its special subalgebra (defined only if $n=0$) are
$$
\fh (2n|m)=\{ D\in \fvect (2n|m):\ L_D\omega_0=0\}\; \text{ and} \;
\fsh (m)=\{H_f\in \fh (0|m): \int f\vvol=0\}.
$$
Its odd analogues are the Lie superalgebra of Leitesian fields introduced in
[L1] and its special subalgebra:
$$
\fle (n)=\{ D\in \fvect (n|n): L_D\omega_1=0\} \; \text{ and} \;
\fsle (n)=\{ D\in \fle (n): \Div D=0\}.
$$

It is not difficult to prove the following isomorphisms (as superspaces): 
$$
\renewcommand{\arraystretch}{1.4}
\begin{array}{r@{\,}c@{\,}lr@{\,}c@{\,}l}
\fk (2n+1|m)&\cong&\Span(K_f: f\in \Cee[t, p, q, \xi]);&\fle
(n)&\cong&\Span(Le_f: f\in \Cee [q, \xi]);\\
\fm (n)&\cong&\Span(M_f: f\in \Cee [\tau, q, \xi]);&
\fh (2n|m)&\cong&\Span(H_f: f\in
\Cee [p, q, \xi]).
\end{array}
$$

\begin{rem*}{Remark} 1) It is obvious that the Lie superalgebras of the
series $\fvect$, $\fsvect$, $\fh$ and $\fpo$ for $n=0$ are finite dimensional.

2) A Lie superalgebra of the series $\fh$ is the quotient of the Lie
superalgebra $\fpo$ modulo the one-dimensional center
$\fz$ generated by constant functions.
Similarly, $\fle$ and $\fsle$ are the quotients of $\fb$ and $\fsb$,
respectively, modulo the one-dimensional (odd) center $\fz$ generated by
constant functions. 
\end{rem*}

Set $\fspo (m)=\{ K_f\in \fpo (0|m):\int fv_\xi=0\}$; clearly, $\fsh (m)=\fspo
(m)/\fz$.

Since
$$
\Div M_f =(-1)^{p(f)}2\left ((1-E)\pderf{f}{\tau} - \sum\limits_{i\leq
n}\frac{\partial^2 f}{\partial q_i \partial\xi_i}\right ), 
$$
it follows that
$$
\fsm (n) = \Span\left (M_f \in \fm (n): (1-E)\pderf{f}{\tau}
=\sum\limits_{i\leq n}\frac{\partial^2 f}{\partial q_i
\partial\xi_i}\right ).
$$

In particular, 
$$
\Div Le_f = (-1)^{p(f)}2\sum\limits_{i\leq n}\frac{\partial^2 f}{\partial 
q_i \partial\xi_i}.
$$
The odd analog of the Laplacian, namely, the operator
$$
\Delta=\sum\limits_{i\leq n}\frac{\partial^2 }{\partial 
q_i \partial\xi_i}
$$
on a periplectic supermanifold appeared in physics under the name of {\it
BRST operator}, cf. [GPS]. The divergence-free vector fields from $\fsle (n)$ are
generated by {\it harmonic} functions, i.e., such that $\Delta(f)=0$.

Lie superalgebras $\fsle (n)$, $\fs\fb (n)$ and $\fsvect (1|n)$
have ideals $\fsle \degree(n)$, $\fs\fb \degree(n)$ and $\fsvect
\degree(n)$ of codimension 1 defined from the exact sequences 
$$ 
\renewcommand{\arraystretch}{1.4}
\begin{matrix}
0\longrightarrow \fsle \degree(n)\longrightarrow \fsle (n)\longrightarrow \Cee\cdot
Le_{\xi_1\dots\xi_n} \longrightarrow 0, \\
0\longrightarrow \fs\fb \degree(n)\longrightarrow \fs\fb (n)\longrightarrow \Cee\cdot
M_{\xi_1\dots\xi_n} \longrightarrow 0, \\
\displaystyle 0\longrightarrow \fsvect \degree(n)\longrightarrow \fsvect (1|n)\longrightarrow 
\Cee \cdot\xi_1\dots\xi_n\pder{t}\longrightarrow 0\end{matrix}
$$

\ssec{0.4. Nonstandard realizations} 
In [LSh] we proved that the
following are all the nonstandard gradings of the Lie superalgebras
indicated. Moreover, the gradings in the series
$\fvect$ induce the gradings in the series $\fsvect$, and $\fsvect\degree$; the
gradings in $\fm$ induce the gradings in $\fsm$, $\fle$, $\fsle$,
$\fsle\degree$,
$\fb$, $\fsb$, $\fsb\degree$; the gradings in $\fk$ induce the gradings in
$\fpo$, $\fh$. In what follows we consider $\fk (2n+1|m)$ as preserving Pfaff
eq. $\alpha_1=0$. 

The standard realizations are marked by $(*)$ and in this case indication to $r=0$ is
omitted; note that (bar several exceptions for small
$m, n$) it corresponds to the case of the minimal codimension of ${\cal L}_0$.   
\small
$$
\renewcommand{\arraystretch}{1.2}
\begin{tabular}{|c|c@{\quad}c|}
\hline
Lie superalgebra & its $\Zee$-grading &\\ 
\hline
$\fvect (n|m; r)$, & $\deg u_i=\deg \xi_j=1$  for any $i, j$ & $(*)$\\ 
\cline{2-3}
$ 0\leq r\leq m$ & $\deg \xi_j=0$ for $1\leq j\leq r;\
\deg u_i=\deg \xi_{r+s}=1$ for any $i, s$ & \\ 
\hline
 & $\deg \tau=2$, $\deg q_i=\deg 
\xi_i=1$  for any $i$ & $(*)$\\ 
\cline{2-3}
$\fm(n; r),\; 0\leq r\leq n$& $\deg \tau=\deg q_i=1$, $\deg 
\xi_i=0$ for any $i$ & \\ 
\cline{2-3}
& $\deg \tau=\deg q_i=2$, $\deg \xi_i=0$ for $1\leq i\leq r
<n$; & \\ 
& $\deg u_{r+j}=\deg \xi_{r+j}=1$ for any $j$ &\\ 
\hline
$\fk (2n+1|m; r)$, & $\deg t=2$, $\deg p_i=\deg q_i=
\deg \xi_j=\deg \eta_j=\deg \theta_k=1$ for any $i, j, k$ & $(*)$ \\ 
\cline{2-3}
$0\leq r\leq [\frac{m}{2}]$ & $\deg t=\deg \xi_i=2$, $\deg 
\eta_{i}=0$ for $1\leq i\leq r\leq [\frac{m}{2}]$; & \\
&$\deg p_i=\deg q_i=\deg \theta_{j}=1$ for
$j\geq 1$ and all $i$ &\\ 
\hline
$\fk(1|2m; m)$ & $\deg t =\deg \xi_i=1$, $\deg 
\eta_{i}=0$ for $1\leq i\leq m$ & \\
\hline
\end{tabular}
$$
\vskip 0.2 cm

\normalsize
Observe that $\fk(1|2; 2)\cong\fvect(1|1)$ and $\fm(1; 1)\cong\fvect(1|1)$.

{\bf The exceptional nonstandard
gradings}. Denote the indeterminates and their respective exceptional degrees as follows
(here
$\fk(1|2)$ is considered in the realization that preserves the Pfaff eq.$\alpha_1=0$): 
$$
\renewcommand{\arraystretch}{1.2}
\begin{tabular}{|r|c|c|l|}
\hline
$\fvect(1|1)$&$t,  \xi$& 2, 1&1, $-1$\\
\hline
$\fk (1|2)$&$t, \xi,  \eta$& $1,  2,  -1$&\\
\hline
$\fm(1)$&$\tau,  q,  \xi$&1,  2,  -1&\\
\hline
\end{tabular}
$$
Denote the nonstandard exceptional realizations by indicating the above degrees after a
semicolon. We
get the following isomorphisms:
$$
\renewcommand{\arraystretch}{1.4}
\begin{array}{cc} 
\fvect (1|1; 2,  1) \cong \fk (1|2);\; \; &\fk (1|2; 1,  2,  -1) \cong \fm (1);\\  
\fvect (1|1; 1,  -1) \cong \fm (1);\; \; & \fm (1; 1,  2,  -1) \cong \fk (1|2). 
\end{array}
$$

Observe that the Lie superalgebras corresponding to different values of $r$ are isomorphic as
abstract Lie superalgebras, but as filtered ones they are distinct.

\ssec{0.5. The Cartan prolongs} 
We will repeatedly use the Cartan prolong. So
let us recall the definition and generalize it somewhat. Let $\fg$ be a Lie
algebra, $V$ a $\fg$-module, $S^i$ the operator of the $i$-th symmetric
power. Set $\fg_{-1} = V$, $\fg_0 = \fg$ 
and define the $i$-th {\it Cartan prolongation} for $i > 0$ as

\begin{multline*}
\fg_i = \{X\in \Hom(\fg_{-1}, \fg_{i-1}): X(v)(w,...) = X(w)(v,...)\;
\text{ for any }\; v, w\in \fg_{-1}\}\\
= (S^i(\fg_{-1})^*\otimes \fg_0)\cap (S^{i+1}(\fg_{-1})^*\otimes \fg_{-1}). 
\end{multline*}

The {\it Cartan prolong} (the result of 
Cartan's {\it prolongation}) of the pair $(V, \fg)$ is $(\fg_{-1},
\fg_{0})_* = \mathop{\oplus}\limits_{i\geq -1} \fg_i$. (In what follows ${\bcdot}$ in
superscript denotes, as is now customary, the collection of all degrees, while $*$ is
reserved for dualization; in the subscripts we retain the oldfashioned $*$ instead of
${\bcdot}$ to avoid too close a contact with the punctuation marks.)
	
Suppose that the $\fg_0$-module $\fg_{-1}$ is faithful. Then, clearly, 

\begin{multline*}
(\fg_{-1}, \fg_{0})_*\subset \fvect (n) = \fder~
\Cee[x_1,..., x_n],\; \text{ where }\; n = dim~ \fg_{-1}\; \text{ and }\\
\fg_i = \{D\in \fvect(n): \deg D=i, [D, X]\in\fg_{i-1}\text{ for any }
X\in\fg_{-1}\}. 
\end{multline*}

It is subject to an easy verification that the Lie algebra structure on
$\fvect (n)$ induces same on $(\fg_{-1}, \fg_{0})_*$. 

Of the four simple vectoral Lie algebras, three are Cartan prolongs:
$\fvect(n)=(\id, \fgl(n))_*$, $\fsvect(n)=(\id, \fsl(n))_*$ and
$\fh(2n)=(\id, \fsp(n))_*$. The fourth one --- $\fk(2n+1)$ --- is also the
prolong under a trifle more general construction described as follows.

\ssec{A generalization of the Cartan prolong} 
Let $\fg_-=\mathop{\oplus}\limits_{-d\leq i\leq -1}\fg_i$ be a nilpotent $\Zee$-graded Lie
algebra and $\fg_0\subset \fder_0\fg$ a Lie subalgebra of the $\Zee$-grading-preserving
derivations. For $i >0$ define the $i$-th prolong of the pair $(\fg_, \fg_0)$
to be: 
$$ 
\fg_i = ((S^{\bcdot}(\fg_-)^*\otimes \fg_0)\cap (S^{\bcdot}(\fg_-)^*\otimes
\fg_-))_i, 
$$ 
where the subscript $i$ in the rhs singles out the component of degree $i$. 

Define $\fg_*$, or rather, $(\fg_-, \fg_0)_*$, to be
$\mathop{\oplus}\limits_{i\geq -d}
\fg_i$; then, as is easy to verify, $(\fg_-, \fg_0)_*$ is a Lie algebra. 

What is the Lie algebra of contact vector fields in these terms? Denote by
$\fhei(2n)$ the Heisenberg Lie algebra: its space is $W\oplus {\Cee}\cdot z$,
where $W$ is a $2n$ dimensional space endowed with a nondegenerate skew-symmetric
bilinear form $B$ and the bracket in $\fhei(2n)$ is given by the
following conditions: $z$ is in the center and $[v, w]=B(v, w)\cdot z$ for any $v, w\in
W$.

Clearly, $\fk(2n+1)$ is
$(\fhei(2n), \fc\fsp(2n))_*$, where for any $\fg$ we write
$\fcg=\fg \oplus {\Cee}\cdot z$ or $\fc(\fg)$ to denote the trivial central
extension with the 1-dimensional even center generated by $z$. 

\ssec{0.6. Lie superalgebras of vector fields as Cartan's prolongs} The
superization of the constructions from sec. 0.5 are straightforward: via Sign
Rule. We thus get $\fvect(m|n)=(\id, \fgl(m|n))_*$, $\fsvect(m|n)=(\id, \fsl(m|n))_*$,
$\fh(2m|n)=(\id, \fosp^{sk}(m|2n))_*$; $\fle(n)=(\id, \fpe^{sk}(n))_*$,
$\fs\fle(n)=(\id, \fspe^{sk}(n))_*$. 

{\it Remark}. Observe that the Cartan prolongs $(\id, \fosp^{sy} (m|2n))_*$
and $(\id, \fpe ^{sy} (n))_*$ are finite dimensional. 

The generalization of Cartan's prolongations described in 0.5 has,
after superization, two analogs associated with the contact series $\fk$ and
$\fm$, respectively. 

$\bullet$ First we define $\fhei(2n|m)$ on the direct sum of a $(2n,
m)$-dimensional superspace $W$ endowed with a nondegenerate skew-symmetric
bilinear form and a $(1, 0)$-dimensional space spanned by $z$.

Clearly, we have $\fk(2n+1|m)=(\fhei(2n|m), \fc\fosp^{sk}(m|2n))_*$ and, given
$\fhei(2n|m)$ and a subalgebra $\fg$ of $\fc\fosp^{sk}(m|2n)$, we call
$(\fhei(2n|m), \fg)_*$ the {\it $k$-prolong} of $(W, \fg)$, where $W$ is the
identity $\fosp^{sk}(m|2n)$-module.

$\bullet$ The odd analog of $\fk$ is associated with the following odd analog of
$\fhei(2n|m)$. Denote by $\fab(n)$ the {\it antibracket} Lie superalgebra: its
space is $W\oplus \Cee\cdot z$, where $W$ is an $n|n$-dimensional superspace
endowed with a nondegenerate skew-symmetric odd bilinear form $B$; the
bracket in $\fab(n)$ is given by the following formulas: $z$ is odd and lies in
the center; $[v, w]=B(v, w)\cdot z$ for $v, w\in W$.

Set $\fm(n)=(\fab(n), \fc\fpe^{sk}(n))_*$ and, given $\fab(n)$ and
a subalgebra $\fg$ of $\fc\fpe^{sk}(n)$, we call $(\fab(n),
\fg)_*$ the {\it $m$-prolong} of $(W, \fg)$, where $W$ is the identity
$\fpe^{sk}(n)$-module.

Generally, given a nondegenerate form $B$ on a superspace $W$ and a
superalgebra $\fg$ that preserves $B$, we refer to the above generalized
prolongations as to {\it $mk$-prolongation} of the pair $(W, \fg)$. 

{\bf A partial Cartan prolong or the prolong of a positive part}. Take a $\fg_0$-submodule
$\fh_1$ in $\fg_1$. Suppose that $[\fg_{-1}, \fh_1]=\fg_0$, not a subalgebra
of $\fg_0$. Define the 2nd prolongation of 
$(\mathop{\oplus}\limits_{i\leq 0}\fg_i, \fh_1)$ to be $\fh_{2}=\{D\in\fg_{2}: [D, \fg_{-1}]\in
\fh_1\}$. The terms $\fh_{i}$ are similarly defined. Set $\fh_i=\fg_i$ for
$i<0$ and $\fh_*=\sum\fh_i$.

{\it Examples}: $\fvect(1|n; n)$ is a subalgebra of
$\fk(1|2n; n)$. The former is obtained as Cartan's prolong of the same nonpositive part
as $\fk(1|2n; n)$ and a submodule of $\fk(1|2n; n)_1$, cf. Table 0.9. The simple exceptional
superalgebra
$\fk\fas$ introduced in 0.9 is another example.

\ssec{0.7. The modules of tensor fields} To advance further, we have to recall the
definition of the modules of tensor fields over $\fvect(m|n)$ and its subalgebras, see [BL].
Let
$\fg=\fvect(m|n)$ (for any other $\Zee$-graded vectoral Lie superalgebra the construction is
identical) and
$\fg_{\geq}=\mathop{\oplus}\limits_{i\geq 0}\fg_{i}$. Clearly, $\fvect_0(m|n)\cong \fgl(m|n)$.
Let $V$ be the $\fgl(m|n)$-module with the {\it lowest} weight $\lambda=\lwt(V)$. Make
$V$ into a $\fg_{\geq}$-module setting $\fg_{+}\cdot V=0$ for
$\fg_{+}=\mathop{\oplus}\limits_{i> 0}\fg_{i}$. Let us realize $\fg$ by vector fields on the
$m|n$-dimensional linear supermanifold $\cC^{m|n}$ with coordinates $x=(u, \xi)$. The
superspace $T(V)=\Hom_{U(\fg_{\geq})}(U(\fg), V)$ is isomorphic, due to the
Poincar\'e--Birkhoff--Witt theorem, to ${\Cee}[[x]]\otimes V$. Its elements
have a natural interpretation as formal {\it tensor fields of type} $V$. When $\lambda=(a, \dots , a)$
we will simply write $T(\vec a)$ instead of $T(\lambda)$. We usually consider irreducible
$\fg_0$-modules.

{\it Examples}: $T(\vec 0)$ is the superspace of functions; $\Vol(m|n)=T(1, \dots , 1;
-1, \dots , -1)$ (the semicolon separates the first $m$ coordinates of the
weight with respect to the matrix units $E_{ii}$ of $\fgl(m|n)$) is the
superspace of {\it densities} or {\it volume forms}. We denote the generator
of $\Vol(m|n)$ corresponding to the ordered set of coordinates $x$ by $\vvol(x)$. The space
of $\lambda$-densities is $\Vol^{\lambda}(m|n)=T(\lambda, \dots , \lambda;
-\lambda, \dots , -\lambda)$. In particular, $\Vol^{\lambda}(m|0)=T(\vec \lambda)$ but
$\Vol^{\lambda}(0|n)=T(\overrightarrow{-\lambda})$.

{\it Remark}. To view the volume element as \lq\lq $d^mud^n\xi$" is totally wrong. Though
under linear transformations that do not intermix the even $u$ with the odd $\xi$ the volume
element $\vvol(x)$ can be viewed as the fraction 
$\frac{du_1\cdot ...\cdot du_m}{d\xi_1\cdot ...\cdot d\xi_n}$. Here {\it all} the
differentials anticommute. How could this happen? If we consider the usual, exterior,
differential forms, then the $d\xi_i$ commute, if we consider the {\it symmetric} product
of the differentials, as in the metrics, then the $dx_i$ commute. However, the
$\pder{\xi_i}$ anticommute and, from transformation point of view,
$\pder{\xi_i}=\frac{1}{d\xi_{i}}$. The notation, $du_1\cdot ...\cdot du_m\cdot\pder{\xi_1}\cdot...\cdot
\pder{\xi_n}$, is, nevertheless, still wrong: almost any transformation $A: (u, \xi)\mapsto
(v, \eta)$ sends $du_1\cdot ...\cdot du_m\cdot\pder{\xi_1}\cdot...\cdot
\pder{\xi_n}$ to the correct element, $\ber (A)(du^m\cdot\pder{\xi_1}\cdot...\cdot
\pder{\xi_n})$, plus extra terms. Indeed, the fraction $du_1\cdot ...\cdot
du_m\cdot\pder{\xi_1}\cdot ...\cdot
\pder{\xi_n}$ is the highest weight vector of an {\it indecomposable}
$\fgl(m|n)$-module and $\vvol(x)$ is the image of this vector in the 1-dimensional quotient
module modulo the invariant submodule that consists precisely of the extra terms. 

\ssec{0.8. Deformations of the Buttin superalgebra}  Here we reproduce a result of Kotchetkov [Ko1]
with corrections from [Ko2], [L3], [LSh]. As is clear from the definition
of the Buttin bracket, there is a regrading (namely, $\fb (n; n)$ given by
$\deg\xi_i=0, \deg q_i=1$ for all $i$) under which $\fb(n)$, initially
of depth 2, takes the form $\fg=\mathop{\oplus}\limits_{i\geq -1}\fg_i$ with $\fg_0=\fvect(0|n)$
and
$\fg_{-1}\cong
\Pi(\Cee[\xi])$.

Let us replace the $\fvect(0|n)$-module $\fg_{-1}$ of functions (with reversed
parity) with the module of $\lambda$-densities, i.e., set $\fg_{-1}\cong
\Cee[\xi](\vvol_\xi)^\lambda$, where 
$$
L_D(vol_\xi)^\lambda
=\lambda \Div D\cdot vol_\xi^\lambda\; \text{ and }\;
p(vol_\xi)^\lambda=\od. 
$$
Then the Cartan's prolong $(\fg_{-1}, \fg_0)_*$ is a deform $\fb_{\lambda}(n;
n)$ of $\fb(n; n)$. The collection of these deforms for various $\lambda\in\Cee$
constitutes a deformation of $\fb(n; n)$; we called it the {\it main
deformation}.\index{deformation of the Buttin bracket, main} (Though main, it
is not the quantization of the Buttin bracket, cf. [L3].) The deform
$\fb_{\lambda}(n)$ of $\fb(n)$ is the regrading of
$\fb_{\lambda}(n; n)$  inverse to the regrading of $\fb(n; n)$ into
$\fb(n)$.

Another description of the main deformation is as follows. Set 
$$
\fb_{a, b}(n) =\{M_f\in \fm (n):\ a\; \Div M_f=(-1)^{p(f)}2(a-bn)\pderf{f}{\tau}\}.
$$

It is subject to a direct check that $\fb_{a, b}(n)\cong \fb_\lambda(n)$ for
$\lambda =\frac{2a}{n(a-b)}$. This isomorphism shows that $\lambda$ actually
runs over $\Cee\Pee^1$, not $\Cee$. Observe that for $a=nb$, i.e., for
$\lambda=\frac{2}{n-1}$, we have $\fb_{nb, b}(n)\cong \fsm(n)$.

As follows from the description of $\fvect(m|n)$-modules ([BL]) and the
criteria for simplicity of $\Zee$-graded Lie superalgebras ([K]),
the Lie superalgebras $\fb_\lambda(n)$ are simple for
$n>1$ and $\lambda\neq 0, \ -1, \infty$. It is also clear that the $\fb_{\lambda}(n)$
are nonisomorphic for distinct $\lambda$'s. (Notice, that at some values of $\lambda$ the
Lie superalgebras $\fb_{\lambda}(n)$ have additional deformations distinct from the above.
These deformations are partly described in [L3].)

\ssec{0.9. The exceptional Lie subsuperalgebra 
$\fk\fas$ of $\fk (1|6)$} The Lie superalgebra $\fg=\fk(1|2n)$ is generated by the
functions from $\Cee[t, \xi_1, \dots, \xi_n, \eta_1, \dots, \eta_n]$. The
standard $\Zee$-grading of $\fg$ is induced by the $\Zee$-grading of
$\Cee[t, \xi, \eta]$ given by $\deg t=2$, $\deg
\xi_i=\deg\eta_i=1$; namely, $\deg K_f=\deg f-2$. Clearly, in this grading $\fg$
is of depth 2. Let us consider the functions that generate several first
homogeneous components of
$\fg=\mathop{\oplus}\limits_{i\geq -2}\fg_i$:
$$
\renewcommand{\arraystretch}{1.3}
\begin{array}{|c|c|c|c|c|}
\hline
\text{component}&\fg_{-2}&\fg_{-1}&\fg_{0}&\fg_{1}\\
\hline
\text{its generators}&1&\Lambda^1(\xi, \eta)&\Lambda^2(\xi,
\eta)\oplus\Cee\cdot t&\Lambda^3(\xi,
\eta)\oplus t\Lambda^1(\xi, \eta)\\
\hline
\end{array}
$$
As one can prove directly, the component $\fg_1$ generates the whole subalgebra
$\fg_+$ of elements of positive degree. The component $\fg_1$ splits into two $\fg_0$-modules
$\fg_{11}= \Lambda^3$ and 
$\fg_{12}=t\Lambda^1$. It is obvious that $\fg_{12}$ is
always irreducible and the component $\fg_{11}$ is
trivial for $n=1$.

Recall that if the operator $d$ that determines a $\Zee$-grading of the Lie superalgebra
$\fg$  does not belong to $\fg$, we denote the Lie
superalgebra $\fg\oplus {\Cee}\cdot d$ by $\fdg$. Recall also that $\fc(\fg)$ or
just $\fc\fg$ denotes the trivial 1-dimensional central extension of $\fg$ with the even
center.

The Cartan prolongations from these components are well-known:
$$
\renewcommand{\arraystretch}{1.4}
\begin{matrix}
(\fg_-\oplus\fg_0, \fg_{11})_*^{mk}\cong\fpo(0|2n)\oplus\Cee\cdot
K_t\cong\fd(\fpo(0|2n));\\
(\fg_-\oplus\fg_0,
\fg_{12})_*^{mk}=\fg_{-2}\oplus\fg_{-1}\oplus\fg_{0}
\oplus\fg_{12}\oplus\Cee\cdot
K_{t^{2}}\cong\fosp(2n|2).\end{matrix}
$$
Observe a remarkable property of $\fk(1|6)$. For $n>1$ and $n\neq 3$ the component
$\fg_{11}$ is irreducible. For $n=3$ it splits into 2 irreducible conjugate
modules that we will denote $\fg_{11}^\xi$ and $\fg_{11}^\eta$. Observe
further, that $\fg_0=\fc\fo(6)\cong\fgl(4)$. As $\fgl(4)$-modules,
$\fg_{11}^\xi$ and $\fg_{11}^\eta$ are the symmetric squares $S^2(\id)$
and $S^2(\id^*)$ of the identity 4-dimensional representation and its dual,
respectively.

\begin{Theorem} The Cartan prolong
$(\fg_-\oplus\fg_0, \fg_{11}^\xi\oplus\fg_{12})_*^{mk}$ is infinite
dimensional and simple. It is isomorphic to $(\fg_-\oplus\fg_0,
\fg_{11}^\eta\oplus\fg_{12})_*^{mk}$.
\end{Theorem}

We will denote the simple exceptional vectoral Lie superalgebra described in Theorem
0.9 by $\fk\fas$.

\ssec{0.10. Several first terms that determine the Cartan and
$mk$-prolongations} To facilitate the comparizon of various
vectoral superalgebras, consider the following Table. The central element $z\in\fg_0$ is
supposed to be chosen so that it acts on $\fg_k$ as $k\cdot\id$. The sign $\supplus$
(resp. $\subplus$) denotes the semidirect sum with the subspace or ideal on the left
(right) of it; $\Lambda (r)=\Cee[\xi_1, \dots, \xi_r]$ is the Grassmann superalgebra of the
elements of degree 0.
$$
\renewcommand{\arraystretch}{1.3}
\begin{tabular}{|c|c|c|c|}
\hline
$\fg$&$\fg_{-2}$&$\fg_{-1}$&$\fg_0$\cr
\hline
\hline
$\fvect(n|m; r)$&$-$&$\id\otimes\Lambda (r)$&$\fgl(n|m-r)\otimes\Lambda
(r)\supplus\fvect(0|r)$\cr  
\hline
\hline
$\fsvect(n|m; r)$&$-$&$\id\otimes\Lambda (r)$&$\fsl(n|m-r)\otimes\Lambda
(r)\supplus\fvect(0|r)$\cr  
\hline
$\fvect(1|m; m)$&$-$&$\Lambda (m)$&$\Lambda (m)\supplus\fvect(0|m)$\cr  
\hline
$\fsvect(1|m; m)$&$-$&$\Vol (0|m)$&$\fvect(0|m)$\cr  
\hline
$\fsvect\degree(1|m; m)$&$-$&$v\in\Vol (0|m):\int v=0$&$\fvect(0|m)$\cr  
\hline
$\fsvect\degree(1|2)$&$-$&$T^0(0)\cong\Lambda
(2)/\Cee\cdot 1$&$\fvect(0|2)\cong\fsl(1|2)$\cr    
\hline
$\fsvect(2|1)$&$-$&$\Pi(T^0(0))$&$\fvect(0|2)\cong\fsl(2|1)$\cr  
\hline
\hline
$\fk(2n+1|m; r)$&$\Lambda(r)$&$\id\otimes\Lambda
(r)$&$\fc\fosp(m-2r|2n)\otimes\Lambda (r)\supplus\fvect(0|r)$\cr 
\hline
$\fh(2n|m; r)$&$\Lambda(r)/\Cee\cdot 1$&$\id\otimes\Lambda
(r)$&$\fosp(m-2r|2n)\otimes\Lambda (r)\supplus\fvect(0|r)$\cr 
\hline
$\fk(1|2m; m)$&$-$&$\Lambda(m)$&$ \Lambda (m)\supplus\fvect(0|m)$\cr 
\hline
$\fk(1|2m+1; m)$&$\Lambda(m)$&$\Pi(\Lambda
(m))$&$\Lambda (m)\supplus\fvect(0|m)$\cr 
\hline
\end{tabular} 
$$
Recall that $\fb_{a, b}(n)\cong \fb_\lambda(n)$ for
$\lambda =\frac{2a}{n(a-b)}$. Recall that $d$ denotes the operator that determines a
$\Zee$-grading of the Lie superalgebra
$\fg$; $\fc(\fg)$ or just $\fc\fg$ denotes the trivial 1-dimensional central extension of
$\fg$ with the even center.
$$
\renewcommand{\arraystretch}{1.3}
\begin{tabular}{|c|c|c|c|}
\hline
$\fg$&$\fg_{-2}$&$\fg_{-1}$&$\fg_0$\cr
\hline
\hline
$\fb_{\lambda}(n;
r)$&$\Pi(\Lambda(r))$&$\id\otimes\Lambda(r)$&$(\fspe(n-r)\supplus\Cee(az+bd))\otimes\Lambda
(r)\supplus\fvect(0|r)$\cr 
\hline
$\fb_{\lambda}(n; n)$&$-$&$\Pi(\Vol^{\lambda}(0|n))$&$\fvect(0|n)$\cr  
\hline
\hline
$\fm(n; r)$&$\Pi(\Lambda(r))$&$\id\otimes\Lambda(r)$&$\fc\fpe(n-r)\otimes\Lambda
(r)\supplus\fvect(0|r)$\cr 
\hline
$\fm(n; n)$&$-$&$\Pi(\Lambda(n))$&$ \Lambda
(n)\supplus\fvect(0|n)$\cr  
\hline
$\fsb\degree_{\lambda}(n; n)$&$-$
&$\frac{\Pi(\Vol(0|n))}{\Cee(1+\lambda\xi_1\dots\xi_n)\vvol(\xi)}$&
$\fsvect_\lambda (0|n)$\cr 
\hline
\hline
$\fle(n; r)$&$\Pi(\Lambda(r))/\Cee\cdot 1$&$\id\otimes\Lambda
(r)$&$\fpe(n-r)\otimes\Lambda (r)\supplus\fvect(0|r)$\cr 
\hline
\hline
$\fle(n; n)$&$-$&$\Pi(\Lambda(n))/\Cee\cdot 1$&
$\fvect (0|n)$\cr 
\hline
$\fsle\degree(n; r)$&$\Pi(\Lambda(r))/\Cee\cdot 1$&$\id\otimes\Lambda
(r)$&$\fspe(n-r)\otimes\Lambda (r)\supplus\fvect(0|r)$\cr 
\hline
$\fsle\degree(n; n)$&$-$&$\Pi(T^0(0))$&$\fsvect(0|n)$\cr 
\hline
\end{tabular} 
$$

\section*{\protect\S 1. Stringy superalgebras}

These superalgebras are particular cases of the Lie superalgebras of vector fileds, 
namely, the ones that preserve a structure on a what physicists call superstring, 
i.e.,  the supermanifold associated with a vector bundle on the circle. These
superalgebras themselves are \lq\lq stringy" indeed: as modules over the Witt
algebra $\fwitt=\fder~\Cee [t^{-1}, t]$ they are direct sums of several \lq\lq
strings", the modules $\cF^\lambda$ described in sec. 2.3. 

This description, sometimes taken for
definition of the stringy superalgebra $\fg$, depends on the embedding $\fwitt\tto\fg$ and the
spectrum of $\fwitt$-modules constituting $\fg$ might vary hampering recognition. Rigorous is a deep
definition of a {\it deep} superalgebra due to Mathieu. He separates the deep
algebras of which stringy is a particular case Lie algebras from affine Kac--Moody ones. Both
are of infinite depth (see 0.2) but while for the loop algebras all root vectors act locally
nilpotently, whereas $\fg$ is {\it stringy} if
$$
\text{$\fg$ is of polynomial growth and there is a root vector which
does not act locally nilpotently}.\eqno{(1.0)}
$$

Similarly, we say that a Lie superalgebra $\fg$ of infinite
depth is of the {\it loop type} if it satisfies (1.0) and {\it stringy} otherwise. Observe,
that a stringy superalgebra of polynomial growth can be a Kac--Moody superalgebra, i.e., have
a Cartan matrix, but not be a (twisted) loop superalgebra. (Roughly speaking, the stringy
superalgebras have the root vector $\frac{d}{dt}$.)

\ssec{1.1} Let $\varphi$ be the angle parameter on the circle, $t= exp(i \varphi)$. The only
stringy Lie algebra is $\fwitt$.

Examples of {\it stringy Lie superalgebras} are certain subalgebras of the Lie superalgebra of
superderivations of either of the two supercommutative superalgebras
$$
R^{L}(1|n)=\Cee [t^{-1}, t, \theta _{1}, \ldots , \theta _{n}]\quad\text{or}\quad R^{M}(1|n)
=\Cee [t^{-1}, t, \theta _{1}, \ldots, \theta _{n-1}, 
\sqrt{t} \xi ].
$$ 

$R^{L}(1|n)$ can be considered as the superalgebra of complex-valued functions expandable into
finite Fourier series or, as superscript indicates,  Laurent series. These functions are
considered on the real supermanifold $S^{1|n}$ associated with the
rank $n$ trivial bundle over the circle. We can forget about $\varphi$ and think in
terms of $t$ considered as the even coordinate on $(\Cee^*)^{1|n}$.

$R^{M}(1|n)$ can be considered as the superalgebra of complex-valued functions (expandable into
finite Fourier series) on the supermanifold $S^{1|n-1, M}$ associated with the Whitney
sum of the M\"obius bundle and the trivial bundle of rank $n-1$. Since the Whitney sum of
two M\" obius bundles is isomorphic to the trivial bundle of rank 2, it suffices to consider
one M\"obius summand.

Let us introduce first stringy Lie superalgebras. These are analogues of
$\fvect$, $\fsvect$ and $\fk$ obtained by replacing $R(1|n) = \Cee [t, \theta _{1}, \ldots, \theta
_{n}]$ with
$R^{L}(1|n)$:
$$
\begin{aligned}
\fvect ^{L}(1|n) &= \fder~ R^{L}(1|n);\\
\fsvect^{L}_\lambda(1|n) &= \{ D\in \fvect ^{L}(1|n)\ :\Div(t^\lambda D) = 0\}=
 \{ D\in \fvect ^{L}(1|n)\ :L_D(t^\lambda\vvol(t, \theta)) = 0\};\\
\fk ^{L}(1|n) &= \{D \in\fvect ^{L}(n): D (\alpha _{1}) = f_ D \alpha _{1}\; 
\text{for}\; \tilde\alpha _{1}=dt + \sum \theta _{i} d\theta _{i}\; \text{and}\; f_D \in
R^{L}(1|n) \}.
\end{aligned}  
$$
We abbreviate $\fsvect^{L}_0(1|n)$ to $\fsvect^{L}(1|n)$.

The routine arguments prove that the elements that the functions $f\in R^{L}(n)$ generate
$\fk^{L}(n)$, and the formulas for $K_{f}$ and 
the contact bracket are the same as for $\fk (1|n)$ and the polynomial $f$.

\begin{rem*}{Exercise} The algebras $\widetilde{\fvect}(1|n)$ and
$\widetilde{\fsvect}_\lambda(1|n)$ obtained by replacing
$R(n)$ with $R^{M}(1|n)$ are isomorphic to ${\fvect}^{L}(1|n)$ and
${\fsvect}^{L}_{\lambda-\frac 12}(1|n)$, respectively. Moreover,
${\fsvect}^{L}_{\lambda}(1|n)\cong {\fsvect}^{L}_{\mu}(1|n)$ if and only if
$\lambda-\mu\in\Zee$. 

The following formula is convenient:
$$
D=f\partial_t+\sum f_i\partial_i\in \fsvect^{L}_{\lambda}(1|n)\quad\text{if and only
if }\quad \lambda f=-t\Div D.\eqno{(1.1.1)}
$$
\end{rem*}

If $\lambda\in\Zee$, the Lie superalgebra ${\fsvect}^{L}_{\lambda}(1|n)$ has the
simple ideal ${\fsvect}^{L0}_{\lambda}(1|n)$ of codimension $\eps^n$:
$$
0\tto {\fsvect}^{L0}_{\lambda}(1|n)\tto {\fsvect}^{L}_{\lambda}(1|n)\tto
\theta_1\cdot\ldots\cdot\theta_n\partial_t\tto 0.
$$ 

$\bullet$ The lift of the contact structure from
$S^{1|n}$ to its two-sheeted covering,
$S^{1|n, M}$, brings a new structure. Indeed, this lift means replacing $\theta_{n}$
with $\sqrt{t} \theta$; this replacement sends the form $\tilde\alpha_{1}$ 
into the M\"obius form  
$$ 
\hat{\tilde\alpha}= dt + \sum_{i=1}^{n-1} \theta_{i} d\theta_{i} +t \theta
d\theta.\eqno{(\widetilde{1.1.2})}
$$ 

It is often convenient to pass to another canonical expression of the {\it M\"obius form}: 
$$
\hat \alpha=\left\{
\begin{array}{ll}
dt + \sum_{i \leq k}(\xi_{i}
d\eta_{i} + \eta_{i}d\xi_{i}) + t
\theta d \theta &\text{if}\; n=2k+1 \\
dt +\sum_{i \leq k}(\xi_{i}
d\eta_{i} + \eta_{i}d\xi_{i} +\zeta d\zeta) + t\theta d \theta & 
\text{if}\;  n=2k+2 
\end{array} \right. \eqno{(1.1.2)}
$$

Now, we have two ways for describing the vector fields that preserve
$\tilde \alpha$ or $\hat{\tilde \alpha}$: 

1) We can set:
$$
{\fk}^{M}(1|n) = \{ D \in \fder \, R^{M}(1|n) \, :\; L_D(\alpha_1) = f_D\cdot
\alpha_{1},  \; \text{ where } \; f_D\in R^{M}(1|n) \}. \eqno{({\faut}_{R^M}(\alpha _{1}))} 
$$
 
In this case the fields $K_{f}$ are given by the same formulas
as for $\fk(1|n)$ but the generating functions belong to
$R^{M}(n)$. The contact bracket between the generating functions from $R^{M}(n)$
is also given by the same formulas as for the generating functions of $\fk(1|n)$.

2) We can set:
$$ 
\fk^{M}(1|n) =\{ D \in \fvect ^{L}(1|n):  L_D (\hat\alpha)
= f_D\cdot \hat\alpha, \; \text{ where }\; f_{D} \in R^{L}(1|n)\}.\eqno{(\faut
_{R^{L}}(\hat\alpha))}
$$
It is not difficult to verify that $\fk^{M}(1|n) =\Span(\tilde K_{f}: f\in R^{L}(1|n))$,
where the {\it M\"obius contact field}\index{M\"obius contact field} is given by the formula
$$  
\tilde K_{f} =\bigtriangleup (f) {\cal D}+ {\cal D}(f)E
+\tilde H_{f},\eqno{(1.1.3)}
$$
in which, as in the case of a cylinder $S^{1, n}$, we set $\bigtriangleup = 2 - E$ and
$E=\sum\limits_{i\leq n-1}
\theta_i\pder{\theta_i}+\theta\pder{\theta}$, but where 
$$
\quad{\cal D} =\pder{t} -\frac{\theta}{2t}\pder{\theta}=\frac 12\tilde K_1
$$ 
and where
$$
\renewcommand{\arraystretch}{1.6}
\begin{matrix}
\displaystyle \tilde H_{f}\; =\; (-1)^{p(f)}\bigg(\sum \;
\pderf{f}{\theta_{i}}
\pder{\theta_{i}}
+\; \frac{1}{t} \pderf{f}{\theta}\pder{\theta}\bigg) \quad\text{in the realization
with form $\hat{\tilde\alpha}$;}\\
\displaystyle \tilde H_{f}\; =\; (-1)^{p(f)}\bigg(\sum \;
\big(\pderf{f}{\xi_{i}}
\pder{\eta_{i}}+\pderf{f}{\eta_{i}}
\pder{\xi_{i}}\big)
+\; \frac{1}{t} \pderf{f}{\theta}\pder{\theta}\bigg) \quad\text{in the realization
with form $\hat{\alpha}$ for $n=2k$,}\\
\displaystyle \tilde H_{f}\; =\; (-1)^{p(f)}\bigg(\sum \;
\big(\pderf{f}{\xi_{i}}
\pder{\eta_{i}}+\pderf{f}{\eta_{i}}
\pder{\xi_{i}}\big)+\pderf{f}{\zeta}
\pder{\zeta}
+\; \frac{1}{t} \pderf{f}{\theta}\pder{\theta}\bigg) \quad\text{in the realization
with form $\hat\alpha$ for $n=2k+1$.}
\end{matrix}
$$ 
The corresponding contact bracket of generating functions will be called the
{\it Ramond bracket}; its explicit form is
$$
\{f, g\}_{R.b.}=\triangle (f)\cD(g) -
\cD(f)\triangle (g)-\{f, g\}_{MP.b.},\eqno{(1.1.4)}
$$
where the {\it M\"obius-Poisson bracket}\index{ M\"obius-Poisson
bracket}\index{bracket M\"obius-Poisson} $\{\cdot , \cdot\}_{MP.b}$ is
$$
\{f, g\}_{MP.b}=(-1)^{p(f)}\left (\sum \;
\pderf{f}{\theta_{i}}
\pderf{g}{\theta_{i}}
+\; \frac{1}{t} \pderf{f}{\theta}\pderf{g}{\theta}\right) \quad\text{in the realization
with form $\hat{\tilde\alpha}$}.\eqno{(1.1.5)}
$$ 

Observe that 
$$
L_{K_{f}}(\alpha_1)=K_1(f)\cdot\alpha_1, \quad
L_{\tilde K_{f}}(\hat\alpha)=\tilde K_1(f)\cdot\hat\alpha.\eqno{(1.1.6)}
$$

\begin{rem*}{Remark} Let us give a relation of the brackets with the dot product on the space of
functions. (More exactly, for the Buttin superalgebra the Lie superalgebra structure is
determined on $\Pi(\cF)$; for $\fk$ it is defined on $\cF_{-2}$, see
sec. 1.3, etc.) This relation is often listed as part of the definition of the Poisson algebra which
is, certainly, pure noncence.
$$
\renewcommand{\arraystretch}{1.4}
\begin{matrix}
\{f, gh\}_{k.b.}=\{f, g\}_{k.b.}h+(-1)^{p(f)p(g)}\{f,
g\}_{k.b.}h+K_1(f)gh;\\
\{f, gh\}_{R.b.}=\{f, g\}_{R.b.}h+(-1)^{p(f)p(g)}\{f,
g\}_{R.b.}h+\tilde K_1(f)gh;\\
\{f, gh\}_{m.b.}=\{f, g\}_{m.b.}h+(-1)^{p(f+1)p(g)}\{f,
g\}_{m.b.}h+M_1(f)gh.
\end{matrix}\eqno{(1.1.7)}
$$
The corresponding formulas for the Poisson and Buttin superalgebras are without the third
term.
\end{rem*}

Explicitely, the embedding $i:\fvect^L(1|n)\tto\fk^L(1|2n)$ is given by
the following formula in which
$\Phi=\sum\limits_{i\leq n}\xi_i\eta_i$:
$$
\renewcommand{\arraystretch}{1.3}
\begin{tabular}{|c|c|}
\hline
$D\in\fvect^L(1|n)$ &the generator of $i(D)$\cr
\hline
$f(\xi)t^m\partial_t$&$(-1)^{p(f)}\frac{1}{2^m}f(\xi)(t-\Phi)^m$,\cr
$f(\xi)t^m\partial_i$&$
(-1)^{p(f)}\frac{1}{2^m}f(\xi)\eta_i(t-\Phi)^{m}$.\cr
\hline 
\end{tabular}\eqno{(1.1.8)}
$$

Clearly,
$\fsvect^L_\lambda(1|n)$ is the subsuperspace of
$\fvect^L(1|n)$ spanned by the expressions
$$
f(\xi)(t-\Phi)^m+\sum\limits_i
f_i(\xi)\eta_i(t-\Phi)^{m-1}\; \text{such that}\;
(\lambda +n)f(\xi)=-\sum\limits_i (-1)^{p(f_{i})}\pderf{f_i}{\xi_i}.\eqno{(1.1.9)}
$$ 

The four series of classical stringy superalgebras are: $\fvect^L(1|n)$,
$\fsvect^L_\lambda(1|n)$, $\fk^{L}(1|n)$ and $\fk^{M}(1|n)$. 

\ssec{1.2. Nonstandard gradings of stringy
superalgebras} The Weisfeiler filtrations of vectoral superalgebras with polynomial or formal
coefficients are determined by the maximal subalgebra of finite codimension not containing ideals
of the whole algebra. The corresponding gradings are natural. We believe that the filtrations
and
$\Zee$-gradings of stringy superalgebras induced by Weisfeiler filtrations of the
corresponding vectoral superalgebras with polynomial coefficients are distinguished but we do
not know how to characterize such
$\Zee$-gradings intrinsically.

The gradings described in sec. 0.4 for $\fvect(1|n)$ and  $\fk(1|n)$ induce gradings of
$\fvect^L(1|n)$ and  $\fk^{L}(1|n)$ and even $\fsvect^L_\lambda(1|n)$ since the
elements of latter have as coeffitients the usual Laurent polynomials. For
$\fk^{M}(1|n)$ the preserved form $\hat\alpha$ should
be homogeneous, so $\deg\theta$ is always equal to $0$. In realization
${\fk}^{M}(1|n)=\faut_{R^{L}}(\hat\alpha)$ these $\Zee$-gradings are as follows (here $(*)$
marks  the \lq\lq standard" grading):  
\noindent 
$$ 
\renewcommand{\arraystretch}{1.3}
\begin{tabular}{|l|l|r|} 
\hline
$\fk ^{M}(1|n)$ & \text{$\deg t=2, \;\; \deg \theta =0, \;
\deg \xi _{i}=1 $ for all $ i\qquad\qquad(*)$}\\ 
\hline
$\fk ^{M}(1|2n; r)$
& $ \deg t=\deg \eta _{1}= \ldots =
\deg \eta _{r}=2 \ \deg \xi _{1} = \ldots = \deg \xi _{r}
=0$ \\ 
$1 \leq r < n $
&$\deg \xi _{r+i} = \deg\eta _{r+i} = \deg \zeta = 1 \; \text{ for all }\ i$\\ 
\hline
$\fk ^{M}(1|2n+1; n) $
& $\deg t = \deg \eta _{1} =
\ldots = \deg \eta _{n} = 1$\\
& $\deg \theta =\deg \xi _{1}
= \ldots = \deg \xi _{n} =0$\\
\hline \end{tabular} 
$$

\ssec{1.3. Modules of tensor fields over stringy superalgebras}
Denote by $T^L(V)= \Cee[t^{-1}, t]\otimes V$ the
$\fvect(1|n)$-module that differs from $T(V)$ by allowing the Laurent polynomials as
coefficients of its elements instead of just polynomials. Clearly,
$T^L(V)$ is a $\fvect^L(1|n)$-module. Define
the {\it twisted with weight $\mu$}\index{tensot fields, twisted} version of
$T^L(V)$ by setting: 
$$
T^L_\mu (V)=\Cee[t^{-1}, t]t^\mu\otimes V.
$$

$\bullet$ {\bf The \lq\lq simplest" modules --- the analogues of the
standard or identity representation of the matrix algebras}.  The simplest
modules over the Lie superalgebras of series $\fvect$ are,
clearly, the modules of $\lambda$-densities, cf. sec. 0.7. These modules are
characterized by the fact that they are of rank 1 over $\cF$, the algebra of
functions. Over stringy superalgebras, we can also twist these modules and consider 
$\Vol^\lambda_\mu$. Observe that for $\mu\not\in\Zee$ this module has only one submodule, the
image of the exterior differential $d$, see [BL], whereas for $\mu\in\Zee$ there is,
additionally, the kernel of the residue:
$$
\Res: \Vol^L \longrightarrow
\Cee \, , \;\;\; f\vvol_{t, \xi} \mapsto \text{ the coefficient of }\
\frac{\xi_{1} \ldots \xi_{n}}{t} \ \text{ in the power series expansion of }\ f.
$$

$\bullet$ Over $\fsvect^L(1|n)$ all the spaces $\Vol ^\lambda$ are, clearly,
isomorphic, since their generator, $\vvol(t, \theta)$, is preserved. So all rank 1 modules over
the module of functions are isomorphic to the module of twisted functions $\cF_\mu$.

Over $\fsvect_\lambda^L(1|n)$, the simplest modules are generated by $t^\lambda\vvol(t,
\theta)$. The submodules of the simplest modules over $\fsvect^L(1|n)$ and
$\fsvect_\lambda^L(1|n)$ are the same as those over $\fvect^L(1|n)$ but if $\mu\in\Zee$ there is,
additionally, the trivial submodule generated by (the $\lambda$-th power of) $\vvol(t, \theta)$
or $t^\lambda\vvol(t, \theta)$, respectively

$\bullet$ Over contact superalgebras $\fk(2n+1|m)$, it is more natural to express
the simplest modules not in terms of $\lambda$-densities but via powers of the form
$\alpha=\alpha_1$:
$$
\cF_\lambda=\cases \cF\alpha^\lambda&\text{ for }n=m=0\\
\cF\alpha^{\lambda/2}&\text{otherwise }.\endcases
$$
Observe that $\Vol ^\lambda\cong\cF_{\lambda(2n+2-m)}$, as
$\fk(2n+1|m)$-modules. In particular, the Lie superalgebra of series $\fk$ does not
distinguish between $\frac{\partial}{\partial t}$ and $\alpha^{-1}$: their
transformation rules are identical. Hence, $\fk(2n+1|m)\cong \cases \cF\alpha^\lambda&\text{ for }n=m=0\\
\cF\alpha^{\lambda/2}&\text{otherwise }.\endcases$

$\bullet$ For $n=0, m=2$ (we take $\alpha =dt-\xi d\eta-\eta d\xi$) there are other
rank 1 modules over $\cF$, the algebra of functions, namely: 
$$
T(\lambda, \nu)_\mu=\cF_{\lambda;\mu}\cdot\left(\frac{d\xi}{d\eta}\right)^{\nu/2}.
$$

$\bullet$ Over $\fk^M$, we should replace the form $\alpha$ with
$\tilde \alpha$ and the definition of the $\fk^L(1|m)$-modules $\cF_{\lambda; \mu}$ should be
replaced with
$$
\cF^M_{\lambda; \mu}=\cases \cF_{\lambda; \mu}(\tilde \alpha)^\lambda&\text{ for }m=1\\
\cF_{\lambda; \mu}(\tilde \alpha)^{\lambda/2}&\text{ for }m>1.\endcases
$$ 

$\bullet$ For $m=3$ and $\alpha =dt-\xi d\eta-\eta d\xi-t\theta d\theta$ there are
other rank 1 modules over the algebra of functions $\cF$, namely: 
$$
T^M(\lambda, \nu)_\mu=\cF^M_{\lambda; \mu}\cdot\left(\frac{d\xi}{d\eta}\right)^{\nu/2}.
$$

\begin{rem*}{Examples} 1) The $\fk(2n+1|m)$-module of volume
forms is $\cF _{2n+2-m}$. In particular, $\fk(2n+1|2n+2) \subset
\fsvect(2n+1|2n+2).$

2) As $\fk^L(1|m)$-module, $\fk^L(1|m)$ is isomorphic to $\cF _{-1}$ for
$m=0$ and $\cF _{-2}$ otherwise. As $\fk^M(1|m)$-module, $\fk^M(1|m)$ is isomorphic to 
$\cF _{-1}$ for $m=1$ and $\cF _{-2}$ otherwise. In particular,
$\fk^L(1|4) \simeq \Vol$ and $\fk^M(1|5) \simeq \Pi(\Vol)$.
\end{rem*}

\ssec{1.4. The four exceptional stringy superalgebras} The \lq\lq status" of these
exceptions is different: the first algebra is a true exception, the second one is an
exceptional regrading; the last two are \lq\lq drop outs" from the series.

{\bf A) $\fk\fas^L$} Certain polynomial functions described out in \S 2 generate
$\fk\fas\subset\fk(1|6)$. Inserting Laurent polynomials in the formulas for the generators of
$\fk\fas$ we get the exceptional stringy superalgebra $\fk\fas^L\subset\fk^L(1|6)$.

{\bf B) ${\fm}^{L}(1)$} On the complexification of $S^{1, 2}$, let $q$ be the even
coordinate, $\tau$ and $\xi$ the odd ones. Set 
$$
{\fm}^{L}(1) =\{D\in {\fvect}^{L}(1|2) :D\alpha_{0}= f_{D}\alpha_{0}, \; \text{where}\; \;
f_{D} \in R^{L}(1|1)\, , \; \alpha_{0} = d \tau + q d \xi + \xi dq \}.
$$

{\bf C, D) $\fk^{L0}(1|4)$ and $\fk^{M0}(1|5)$} It follows from Example 2) in sec. 2.3 that
the functions with zero residue on $S^{1|4}$ (resp. $S^{1|4; M}$) generate an ideal in
$\fk^{L}(4)$ (resp.
$\fk^{M}(5)$). These ideals are, clearly, simple Lie superalgebras
denoted in what follows by $\fk^{L0}(1|4)$ and $\fk^{M0}(1|5)$,
respectively.

\ssec{1.5. Deformations} The superalgebrbas ${\fsvect}(1|n)$ and
${\fsvect}^{0}(n)$ do not have deformations that preserve the $\Zee$-gradings
(other deformations may happen; one has to calculate, this is a research
problem). The stringy superalgebras ${\fsvect}^{L}(n)$ do have $\Zee$-grading preserving deformations
discovered by Schwimmer and Seiberg [SS]. More deformations (none of which preserves
$\Zee$-grading) are described in [KvL]; the complete description of deformations of
${\fsvect}^{L}(n)$ is, nevertheless, unknown.

\begin{rem*}{Conjecture} $\fvect^{L}(1|n)$, $\fk^{L}(1|n)$ and the four exceptional stringy
superalgebras are rigid.
\end{rem*} 

\ssec{1.6. Distinguished stringy superalgebras} 

\begin{Theorem} The only nontrivial central extensions of the simple stringy Lie
superalgebras are those given in the following table. 
\end{Theorem} 

Let in this subsection and in sec. 2.1 $K_f$ be the common term for both $K_f$ and
$\tilde K_f$. Let further $\cK = (2\theta\pder{\theta} - 1)\pder{x^2}^2$.
$$ 
\renewcommand{\arraystretch}{1.3}
\begin{tabular}{|c|c|c|}
\hline
algebra & the cocycle $c$ & The name of the extended algebra \\
\hline
$\fk^{L}(1|0)$ & $K_{f}, K_{g}\mapsto \Res fK_{1}^3(g)$ & Virasoro or $\; \fvir$ \\
\hline
$\left. \begin{matrix}\fk^{L}(1|1) \\ 
\fk^{M}(1|1) \end{matrix}\right\}$
 & $K_{f}, K_{g}\mapsto \Res fK_{\theta }K_{1}^2(g)$
 & $\begin{matrix} \text{Neveu-Schwarz or} \; \fns \\
    \text{Ramond or}\; \fr \end{matrix}$ \\
\hline
$\fvect ^{L}(1|1)$ & $D_1=f\pder{t}+g\pder{ \xi}, \;
D_2=\tilde{f}\pder{ t}+\tilde{g}\pder{ \xi}\mapsto\;$& \\  
&$\Res (f\cK(\tilde g) + (-1){p(D_1)p(D_2)}g\cK(\tilde f) +$&$\widehat{\fvect}^{L}(1|1)$\\
&$2(-1){p(D_1)p(D_2)+p(D_2)}g\pder{\theta}\pder{\theta}(\tilde
g)$&\\
\hline
$ \left.\begin{matrix}\fk^{L}(1|2)\\ 
\fk^{M}(1|2) \end{matrix}\right\}$
 & $K_{f}, K_{g}\mapsto \Res fK_{\theta_1 }K_{\theta_2}K_{1}(g)$
 & $\begin{matrix}\text{2-Neveu-Schwarz or}\;\fns(2) \\ 
\text{2-Ramond or}\;\fr(2)\end{matrix}$ \\
\hline
$\fm^{L}(1)$
 & $M_{f}, M_{g}\mapsto \Res f(M_{q})^3(g)$
 & $\widehat{\fm^{L}(1)}$ \\
\hline
$\left.\begin{matrix}\fk  ^{L}(1|3) \\ 
\fk^{M}(1|3)\end{matrix} \right\}$
 & $K_{f}, K_{g}\mapsto \Res fK_{\xi }K_{\theta }K_{\eta }(g)$
 & $\begin{matrix}\text{3-Neveu-Schwarz or}\; \fns(3) \\ 
\text{3-Ramond or}\; \fr (3)\end{matrix}$ \\
\hline
$\left. \begin{matrix}\fk^{L\circ}(4) \\ 
\fk^{M}(1|4)
\end{matrix}\right\}$ & $K_{f}, K_{g}\mapsto \begin{matrix}(1) &
\Res  fK_{\theta_{1}}K_{\theta_{2}}K_{\theta_{3}}K_{\theta _{4}}K_{1}^{-1}(g) \\ 
   (2) & \Res f(tK_{t^{-1}}(g)) \\ 
   (3) & \Res fK_{1}(g)\end{matrix}$
 & $\begin{matrix}(1)\left\{\begin{matrix}\text{4-Neveu-Schwarz}=\fns(4)\\  
\text{4-Ramond}=\fns(4)\end{matrix}\right.\\  
   (2)\left\{\begin{matrix}\text{$4'$-Neveu-Schwarz}=\fns(4')\\  
\text{$4'$-Ramond}=\fns(4')\end{matrix}\right.\\
    (3)\left\{\begin{matrix}\text{$4^0$-Neveu-Schwarz}=\fns(4^0)\\  
\text{$4^0$-Ramond}=\fns(4^0)\end{matrix}\right.\end{matrix}$ \\ 
\hline 
 \end{tabular}   
$$
$$ 
\renewcommand{\arraystretch}{1.3}
\begin{tabular}{|c|c|c|}
\hline
& the restriction of the above cocycle (3): & \\
$\fvect ^{L}(1|2)$ & $D_1=f\pder{t}+g_{1}\pder{\xi
_{1}}+g_{2} \pder{ \xi _{2}}, \;
D_2=\tilde{f}\pder{ t}+\tilde{g}_{1}\pder{ \xi
_{1}}+\tilde{g}_{2}\pder{
\xi_{2}}$& $\widehat{\fvect}^{L}(1|2)$ \\  
&$\mapsto\;\Res(g_{1}\tilde{g}_{2}'-g_{2}\tilde{g}_{1}'(-1)^{p(D_{1})p(D_{2})})$&\\
\hline
$\fsvect ^{L}_{\lambda }(1| 2)$& the restriction of the above  
& $\widehat{\fsvect}^{L}_{\lambda}(1|2)$ \\
\hline   \end{tabular}   
$$

Observe that the restriction of the only cocycle on
$\fvect^{L}(1| 2)$ to its subalgebra $\Span(f(t)\pder{t})\cong\fwitt$ is trivial
while the the restriction of the only cocycle on $\fsvect ^{L}_{\lambda }(1|2)$ to its unique
subalgebra $\fwitt$ is nontrivial. The riddle is solved by a closer study of the embedding
$\fvect(1|m)\tto\fk(1|2m)$: it involves differentiations, see formulas (1.1.8). 

The nonzero values of the cocycle $c$ on $\fvect^{L}(1| 2)$ in monomial basis are:
$$
\renewcommand{\arraystretch}{1.7}
\begin{matrix} 
\displaystyle c(t^k\theta_1\pder{\theta_1}, t^l\theta_2\pder{\theta_2})=k\delta_{k, -l},&
\displaystyle c(t^k\theta_1\pder{\theta_2}, t^l\theta_2\pder{\theta_1})=-k\delta_{k, -l},\\
\displaystyle c(t^k\theta_1\theta_2\pder{\theta_1}, t^l\pder{\theta_2})=k\delta_{k, -l},&
\displaystyle c(t^k\theta_1\theta_2\pder{\theta_2}, t^l\pder{\theta_1})=k\delta_{k, -l}.
\end{matrix} 
$$

In $\fsvect ^{L}_{\lambda}(1|2)$, set:
$$
\renewcommand{\arraystretch}{1.6}
\begin{matrix} 
\displaystyle L_m=t^m\left
(t\pder{t}+\frac{\lambda+m+1}{2}(\theta_1\pder{\theta_1}+\theta_2\pder{\theta_2})\right),\\
\displaystyle S_m^j=t^m\theta_j\left
(t\pder{t}+(\lambda+m+1)(\theta_1\pder{\theta_1}+\theta_2\pder{\theta_2})\right).
\end{matrix} 
$$

The nonzero values of the cocycles on $\fsvect ^{L}_{\lambda }(1|2)$ are
$$
\renewcommand{\arraystretch}{1.6}
\begin{matrix} 
\displaystyle c(L_m, L_n)=\frac 12m(m^2-(\lambda+1)^2)\delta_{m, -n},&
\displaystyle c(t^k\pder{\theta_i}, S_m^j)=-m(m-(\lambda+1))\delta_{m, -n}\delta_{i, j},\\
\displaystyle c(t^m(\theta_1\pder{\theta_1}-\theta_2\pder{\theta_2}),
t^n(\theta_1\pder{\theta_1}-\theta_2\pder{\theta_2}))=m\delta_{m, -n},&
\displaystyle c(t^m\theta_1\pder{\theta_2}, t^n\theta_2\pder{\theta_1})=m\delta_{m, -n}.
\end{matrix} 
$$

When are nontrivial central extensions of
the stringy superalgebras possible? We find the following quantitative discussion
instructive, though it neither replaces the detailed proof (that can be found in [KvL] for
all cases except $\fk\fas^L$; the arguments in the latter case are similar) nor explains the
number of nontrivial cocycles on
$(1|4)$-dimensional supercircle with a contact structure.

When we pass from simple finite dimensional Lie algebras to loop
algebras, we enlarge the maximal toral subalgebra of the latter to make the number of
generators of weight 0 equal to that of positive or negative generators corresponding to
simple roots. In this way we get the nontrivial central extensions of the loop
algebras called {\it Kac--Moody} algebras. (Actually, the latter have one more operator of
weight 0: the exterior derivation.)

Similarly,  for the Witt algebra $\fwitt$ we get:
$$
\renewcommand{\arraystretch}{1.2}
\begin{tabular}{|c||c|c|c|c|c|c|c} 
$\deg K_{f}$&\dots&$-2$&$ -1$&$0$&$1$&$2$&\dots\\
\hline
$f$&\dots&$t^{-1}$&1&$t$&$t^{2}$&$t^{3}$&\dots\\
\end{tabular} 
$$
The depicted elements generate $\fwitt$; more exactly,

(a) the elements of degrees $-1$, $0$, $1$ generate $\fsl (2)$;

(b) $\fwitt$,  as $\fsl(2)$-module, is glued from the three modules:
the adjoint module and the Verma modules $M^{-2}$, and $M_{2}$ with highest and
lowest weights as indicated: $-2$ and 2,  respectively; 

\noindent it is natural to expect a central element to be obtained by pairing of
the dual $\fsl(2)$-modules $M^{-2}$ and $M_{2}$.

This actually happens; one of the methods to find the cocycle is to compute 
$H^{1}(\fg;\,  \fg^{*})$ and identify it with $H^{2}(\fg)$. (The former and the latter
cohomology is what is calculated in [P] and [Sc], respectively.) 

$\bullet$ Further on,  consider the subalgebra $\fosp (n|2)$ in $\fk^{L}(1| n)$ and
decompose $\fk ^{L}(1| n)$, as $\fosp (n|2)$-module, into irreducible
modules. Denote by $(\chi _{0};\, \chi _{1}, \dots , \chi _{r})$ the heighest
(lowest) weight of the irreducible $\fosp (n|2)$-module with respect to the
standard basis of Cartan subalgebra of $\fsl (2)\oplus \fo(n)$; here $r =
[n/2]$.

These modules and their generators are as follows. Set 
$$
\alpha = \left\{
\begin{matrix}
2t-\sum (\xi
_{i}d\eta _{i}+\eta _{i}d\xi _{i})&\text{ if }\;\, n \;\, \text{is even}\\
2t-\sum (\xi
_{i}d\eta _{i}+\eta _{i}d\xi _{i})-\theta d\theta& \text{if}\;\, n\;\,  \text{is odd}\end{matrix}\right.\;\text{ and
}\;\,  \zeta = \left\{\begin{matrix}(\xi , \eta)&\text{ if }\;\, n \text{ is
even }\\ 
(\xi , \eta , \theta )&\text{ if }\;\, n\text{ is odd}.\end{matrix}\right.
$$
Let $\langle f \rangle$  be a shorthand for the generator 
$K_{f}$ of the Verma module $M$ with the highest (lowest) weight as indicated by the
sub- or superscript, respectively; we denote the quotient of $M$ modulo the maximal
submodule by $L$ with the same indices. Then the Jordan--H\"older factors of the
$\fosp(n|2)$-module $\fk^L(1|n)$ are 
$$
\renewcommand{\arraystretch}{1.3}
\begin{matrix}
n&\text{irreducible factors}&\text{of $\fk^L(1|n)$}&\text{as $\fosp (n|2)$-module}\\
0 & \langle t^{-1}\rangle = M^{-2}, &\fsl (2), & M_{2}= \langle t^{3}\rangle \\
1 & \langle t^{-1}\theta \rangle = M^{-3}, &\fosp   
(1| 2), & M_{3} = \langle t^{2}\theta \rangle \\
2 & \langle t^{-1}\xi \eta \rangle = M^{-2;\, 0}, & \fosp(2|2), &
M_{2;\, 0} = \langle t\xi \eta \rangle \\
3 & \langle t^{-1}\xi \eta \theta \rangle = M^{-1;\, 0}, &\fosp (3| 2), & 
M_{1;\, 0} = \langle \xi \eta \theta \rangle \\
4 & \langle t^{-1}\xi _{1}\xi _{2}\eta _{2}\rangle = M^{-1;\, \varepsilon
_{1}}\subplus L_{0;0}=\langle t^{-1}\xi _{1}\xi _{2}\eta _{1}\eta
_{2}\rangle  &\fosp (4|2), & M_{1;\, -\eps _{2}} =
\langle \xi _{1}\eta _{1}\eta _{2}\rangle, \\
5 & \langle t^{-1}\zeta _{1}\dots\zeta _{n}\rangle = M^{1;\, 0}, &
\fosp(5|2), & M_{1;\, -\varepsilon _{1}-\varepsilon _{2}} = \langle \eta _{1}
\eta _{2}\theta \rangle \\
>5 & \langle t^{-1}\zeta _{1}\dots\zeta _{n}\rangle = M^{n-4;\, 0},&
\fosp(n|2),  & M_{1;\, -\varepsilon _{1}-\varepsilon _{2}-\varepsilon _{3}}
= \langle \eta _{1} \eta _{2}\eta _{3}\rangle \end{matrix}
$$
For $n>5$ the module $M_{-1, -\varepsilon _{1}-\varepsilon _{2}-\varepsilon _{3}}$ is
always irreducible whereas $M^{n-4;\, 0}$ is always reducible:
$$
[M^{n-4;\, 0}] = [M^{n-7;\, \varepsilon _{1}+\varepsilon _{2}+\varepsilon_{3}}] \subplus
[L^{n-4;\, 0}].
$$
Exceptional cases: 

$n = 4$. In this case the Verma
module $M^{0;\, 0}$ induced from the Borel subalgebra has an irreducible
submodule $M^{-1;\, \varepsilon _{1}}$ dual to $M_{1;\, -\eps _{2}}$; the
subspace of
$\fk^{L}(1| 4)$ spanned by all functions except $t^{-1}\xi
_{1}\xi _{2}\eta _{1}\eta _{2}$ is an ideal. An explanation of this
phenomenon is given in 1.3. 

$n = 6$. The following miracle happens:
$M^{6-7; \; \varepsilon _{1}+\varepsilon _{2}+\varepsilon _{3}}=
(M_{1; \; -(\varepsilon_{1}+\varepsilon _{2}+\varepsilon _{3})})^{*}$ and
$L^{6-4;\, 0} \cong \fosp (6| 2)$. But the bilinear form obtained is supersymmetric, see \S
2.

For $n>6$ there is no chance to have a nondegenerate bilinear form on
$\fk ^{L}(1| n)$. The above arguments,  however,  do not exclude a degenerate
form on $\fk^{L}(1| n)$ such as a cocycle. For a proof see [KvL].

\ssec{1.7. Root systems and simple roots for $\fsvect^L_\lambda(1|2)$}
Set $\partial=\pder{t}$, $\delta_1=\pder{\xi_1}$,
$\delta_2=\pder{\xi_2}$. 

The generators of the {\it distinguished} system
of simple roots are:
$$
\renewcommand{\arraystretch}{1.4}
\begin{matrix}
X_1^+= \xi_1\delta_2&X_2^+=t\delta_1&X_3^+=\xi_2t\partial-(\lambda+1)\xi_1\xi_2\delta_1\\
X_1^-=\xi_2\delta_1&
\displaystyle X_2^-=\lambda\frac{\xi_1\xi_2}{t}\delta_2+\xi_1\partial&
X_3^-=\delta_2\\
H_1=\xi_1\delta_1-\xi_2\delta_2&H_2=t\partial+\xi_1\delta_1+\lambda\xi_2\delta_2&H_3=
t\partial+(\lambda+1)\xi_1\delta_1
\end{matrix} \eqno{(G1)}
$$
The reflection in the 2nd root sends $(G1)$ into the following system that, to simplify the
expressions, we consider them up to factors in square brackets $[\cdot]$.
$$
\renewcommand{\arraystretch}{1.4}
\begin{matrix}
X_1^+=t\delta_2&
\displaystyle X_2^+=\lambda\frac{\xi_1\xi_2}{t}\delta_2+\xi_1\partial&
X_3^+=[-\lambda]t \xi_2\delta_1\\
\displaystyle X_1^-=\lambda\frac{\xi_1\xi_2}{t}\delta_1-\xi_2\partial&
X_2^-=t\delta_1&
\displaystyle X_3^-= [-\lambda]\frac{\xi_1}{t}\delta_2\\
H_1=-(t\partial+\lambda\xi_1\delta_1+\xi_2\delta_2)&H_2=t\partial+\xi_1\delta_1+
\lambda\xi_2\delta_2&H_3= \xi_2\delta_2-\xi_1\delta_1
\end{matrix} \eqno{(G2)}
$$
The reflection in the 3rd root sends $(G1)$ into the following system. To simplify the
expressions we consider them up to factors in square brackets $[\cdot]$.
$$
\renewcommand{\arraystretch}{1.4}
\begin{matrix}
X_1^+=\delta_2&X_2^+=[\lambda+2]t\xi_2\delta_1&X_3^+=(\lambda+1)
\xi_2\xi_1\delta_2-\xi_1t\partial\\
X_1^-=t\xi_2\partial-(\lambda+1)\xi_1\xi_2\delta_1&
\displaystyle X_2^-=[-\lambda]\frac{\xi_1}{t}\delta_2
&X_3^-= \delta_1\\
H_1=t\partial+(\lambda+1)\xi_1\delta_1&H_2=\xi_2\delta_2-\xi_1\delta_1
&H_3= -t\partial-(\lambda+1)\xi_2\delta_2
\end{matrix} \eqno{(G3)}
$$
The corresponding Cartan matrices are:
$$
\begin{pmatrix} 2&-1&-1\\ 1-\lambda&0&\lambda\\ 1+\lambda&-\lambda&0 \end{pmatrix},\quad
\begin{pmatrix} 0&-\lambda+1&-2+\lambda\\ 1-\lambda&0&\lambda\\ -1&-1&2
\end{pmatrix},\quad
\begin{pmatrix} 0&-\lambda&\lambda+1\\ -1&2&-1\\ 1+\lambda&-\lambda-2&0
\end{pmatrix}.
$$
To compare these matrices, let us reduce them to the following canonical forms $(C1)-(C3)$,
respectively, by renumbering generators and rescaling. (Observe that by definition,
$\lambda \neq 0, \pm 1$, so the fractions are well-defined.) We get
$$
\begin{pmatrix} 2&-1&-1\\-1+\frac{1}{\lambda}&0&1\\1+\frac{1}{\lambda}&-1&0
\end{pmatrix},\quad
\begin{pmatrix} 2&-1&-1\\-1+\frac{1}{1-\lambda}&0&1\\1+\frac{1}{1-\lambda}&-1&0
\end{pmatrix},\quad
\begin{pmatrix} 2&-1&-1\\-1+\frac{1}{1+\lambda}&0&1\\1+\frac{1}{1+\lambda}&-1&0
\end{pmatrix}.
$$
We see that the transformations $\lambda \mapsto \lambda+1$ and $\lambda \mapsto 1-\lambda$
establish isomorphisms. We may, therefore, assume that $\RE \lambda \in[0, \frac{1}{2}]$.

\ssec{1.8. Simplicity and occasional isomorphisms}

\begin{rem*}{Statement} 1) The Lie superalgebras $\fvect^{L}(1|n)$ for any $n$, 
$\fsvect_{\lambda}^{L}(1|n)$ for $\lambda \not\in\Zee$ and $n>1$, ${\fsvect}^{L
0}(1|n)$ for $n>1$, ${\fk}^{M}(1|n)$ for $n \neq 5$ and ${\fk}^{L}(1|n)$ for $n \neq 4$; and the
five exceptional stringy superalgebras are simple.

2) The Lie superalgebras $\fvect ^{L}(1|1)$, $\fk^{L}(1|2)$ and $\fm^{L}(1|1)$ are
isomorphic. 

3) The Lie superalgebras $\fvect^{L}_\lambda(1|2)\cong\fvect^{L}_\mu(1|2)$ if $\mu$ can be obtained
from $\lambda$ with the help of transformations $\lambda \mapsto \lambda+1$ and $\lambda
\mapsto 1-\lambda$. The Lie superalgebras $\fsvect^L_\lambda(1|2)$ from the strip $\RE \lambda
\in [0, \frac{1}{2}]$ are nonisomorphic.
\end{rem*} 

The statement on simplicity follows from a criterion similar to the one Kac applied for Lie
(super)algebras with polynomial coefficients ([K]). The isomorphism is determined by the
gradings listed in sec. 0.4 and arguments of 1.7. 

\ssec{1.9. Remarks. 1) A relation with Kac--Moody superalgebras} An unpublished theorem of
Serganova states that {\sl the only Kac--Moody superalgebras
$\fg(A)$ of polynomial growth with nonsymmetrizable Cartan matrix $A$ are:
$\fp\fsq(n)^{(2)}$ {\em (the corresponding Dynkin--Kac diagram is the same as that of
$\fsl(n)^{(1)}$ but with any odd number of nodes replaced with \lq\lq grey" nodes
corresponding to the odd simple roots, see [FLS])} and an exceptional parametric family
{\em (found by J. van de Leur around 1986)} with the matrix}
$$
A=\begin{pmatrix} 
2 & -1 &-1\\ 
1-\alpha & 0 &\alpha\\
1+\alpha&-\alpha&0
\end{pmatrix}.
$$
The Lie superalgebra $\fg(A)$ can be realized as the distinguished stringy superalgebra
$\widehat{\fsvect}^{L}_{\lambda}(1|2)$. For the description of the relations between its
generators see [GL1].

Observe, that unlike the Kac--Moody superalgebras of polynomial growth with symmetrizable
Cartan matrix, $\hat\fg=\widehat{\fsvect}^{L}_{\lambda}(1|2)$ can not be interpreted as a
central extension of any twisted loop algebra. Indeed, the root vectors of the latter are
locally nilpotent, whereas the former contains the operator $\partial_t$ with nonzero image
of every $\hat\fg_i$.

{\bf 2) How conformal are stringy superalgebras}. Recall that a Lie algebra is called {\it conformal}
if it preserves a metric (or, more generally a bilinear form) $B$ up to a factor. It is known that
given a metric $B$ on the {\it real} space of dimension $\neq 2$, the algebras conformal wrt $B$ are
isomorphic to $\fc(\faut(B))\cong\fc\fo(V, B)$. If $\dim V=2$ we can consider $V$ as the complex line
$\Cee^1$ with coordinate $t$ and identify $B$ with $dt\cdot d\bar t$ (the symmetric product of the
differentials). The element $f\frac{d}{dt}$ from $\fwitt$ multiplies
$dt$ by $f'$ and, therefore, it multiplies $dt\cdot d\bar t$ by  $f'\bar f'$, so
$\fwitt$ is {\it conformal}. 

On superspaces $V$, metrics $B$ can be even and odd, the Lie superalgebras $\faut(V, B)$ that
preserve them are $\fosp^{sy}(\Par)$ and $\fpe^{sy}(\Par)$ and the corresponding conformal
superalgebras are just trivial central extensions of $\faut(V, B)$ for any dimension. 

Suppose now that we consider a real form of each of the stringy superalgebras considered
above and an extension of the complex conjugation (for possibilities see [M]) is defined
in the superspace of the generating functinos. Let the contact superalgebras
$\fk^L$, $\fk^M$ and $\fm^L$ preserve the Pfaff equation with form $\alpha$. From formulas (0.3) and
(1.1.6) we deduce that  the elements of these superalgebras multiply the symmetric product of 
forms $\alpha\cdot\bar\alpha$ by a factor of the form $F\bar F$, where $F$ is the function
determined in (0.3) and (1.1.6). Every element $D$ of the general and divergence-free
superalgebra $\fsvect^L_\lambda$ multiplies the symmetric product of volume forms $\vvol(t,
\theta)\cdot\vvol(\bar t, \bar\theta)$ by $\Div D\cdot\overline{\Div D}$. 

None of the tensors considered can be viewed as a metric. The series $\fsvect^L$
($\lambda\in \Zee$) can not be considered as a superconformal even in the above sence.

There is, however, a possibility to consider  $\fk^L(1|1)$ and  $\fk^M(1|1)$ as conformal
superalgebras, since the volume form \lq\lq $dt\pder{\theta}$" can be considered as, more
or less, $d\theta$: consider the quotient $\Omega^1/\cF\alpha$ of the superspace of
differential 1-forms modulo the subspace spanned over functions by the contact form.
Therefore, the tensor $\displaystyle dt\pder{\theta}\cdot d\bar t\pder{\bar \theta}$ can be
viewed as the bilinear form $d\theta\cdot d\bar\theta$.

\section*{\protect \S 2. Invariant bilinear forms
on stringy Lie superalgebras}

\begin{rem*}{Statement} An invariant (with respect to the adjoint action)
nondegenerate supersymmetric bilinear form on a simple  Lie superalgebra $\fg$, if
exists, is unique up to proportionality. 
\end{rem*}

Observe that the form can be odd. 

The invariant nondegenerate bilinear form
$(\cdot, \cdot)$ on $\fg$ exists if and only if $\fg\cong\left\{\begin{matrix}\fg^*& \text{if
$(\cdot, \cdot)$ is even}\\
\Pi(\fg^*)& \text{if $(\cdot, \cdot)$ is odd}\end{matrix}\right .$ as $\fg$-modules.
Therefore, let us compair $\fg$ with $\fg^*$.  Recall the definition of the modules
$\cF_\lambda$ from sec. 1.3.

$$
\renewcommand{\arraystretch}{1.2}
\begin{tabular}{|c|c|c|c|c|c|c|c|c|c|}
\hline
&0&1&2&3&4&5&6&7&$n$\\
\hline 
$\fg$&$\cF_{-1}$&$\cF_{-2}$&$\cF_{-2}$&$\cF_{-2}$&
$\cF_{-2}$&$\cF_{-2}$&$\cF_{-2}$&$\cF_{-2}$&$\cF_{-2}$\\
\hline 
$\Vol$&$\cF_{1}$&$\Pi(\cF_{0})$&$\cF_{-1}$&$\Pi(\cF_{-2})$&
$\cF_{-3}$&$\Pi(\cF_{-4})$&$\cF_{-5}$&$\Pi(\cF_{-6})$&$\Pi^n(\cF_{2-n})$\\
\hline 
$\fg^*$&$\cF_{2}$&$\Pi(\cF_{3})$&$\cF_{2}$&$\Pi(\cF_{1})$&
$\cF_{0}$&$\Pi(\cF_{-1})$&$\cF_{-2}$&$\Pi(\cF_{-3})$&$\Pi^n(\cF_{4-n})$\\ 
\hline 
\end{tabular}\eqno{(\fg=\fk^L(1|n))}
$$

$$
\renewcommand{\arraystretch}{1.2}
\begin{tabular}{|c|c|c|c|c|c|c|c|c|c|}
\hline
&1&2&3&4&5&6&7&$n$\\
\hline 
$\fg$&$\cF_{-1}$&$\cF_{-2}$&$\cF_{-2}$&$\cF_{-2}$&
$\cF_{-2}$&$\cF_{-2}$&$\cF_{-2}$&$\cF_{-2}$\\
\hline 
$\Vol$&$\Pi(\cF_{1})$&$\cF_{1}$&$\Pi(\cF_{0})$&
$\cF_{-1}$&$\Pi(\cF_{-2})$&$\cF_{-3}$&$\Pi(\cF_{-4})$&$\Pi^n(\cF_{3-n})$\\
\hline 
$\fg^*$&$\Pi(\cF_{2})$&$\cF_{3}$&$\Pi(\cF_{2})$&
$\cF_{1}$&$\Pi(\cF_{0})$&$\cF_{-1}$&$\Pi(\cF_{-2})$&$\Pi^n(\cF_{5-n})$\\ 
\hline 
\end{tabular}\eqno{(\fg=\fk^M(1|n))}
$$
A comparison of $\fg$ with $\fg^*$ shows that there is a nondegenerate bilinear form on
$\fg=\fk^L(1|6)$ and
$\fg=\fk^M(1|7)$, even and odd, respectively. These forms are symmetric and given by the
formula
$$
(K_f, K_g)=\Res fg.
$$
The restriction of the bilinear form to $\fk\fas^L$ is identically zero.

\section*{\S 3. The three cocycles on $\fk^{L0}(1|4)$ and primary fields}
Set 
$$
\renewcommand{\arraystretch}{1.4}
\begin{tabular}{|c|c|c|}
\hline
the elements&their degree &their parity\cr
\hline
$L_n=K_{t^{n+1}};\; T^{ij}_n=K_{t^n\theta_i\theta_j};\;
S_n=K_{t^{n-1}\theta_1\theta_2\theta_3\theta_4}$&$2n$&$\ev$;\cr
$E^i_n=K_{t^{n+1}\theta_i};\; F^{i}_n=K_{t^n\pderf{\theta_1
\theta_2\theta_3\theta_4}{\theta_i}}$&$2n+1$&$\od$.\cr
\hline 
\end{tabular}
$$
In the above \lq\lq natural" basis the nonzero values of the cocycles are (see [KvL];
here $A_n$ is the group of even permutations):
$$
\renewcommand{\arraystretch}{1.4}
\begin{tabular}{|c|}
\hline
$c(L_m, L_n)=\alpha\cdot m(m^2-1)\delta_{m+n, 0}$\cr
$c(E^i_m, E^i_n)=\alpha\cdot m(m+1)\delta_{m+n+1, 0}$\cr
$c(T^{ij}_m, T^{ij}_n)=\alpha\cdot m\delta_{m+n, 0}$\cr
$c(F^i_m, F^i_n)=\alpha\cdot \delta_{m+n+1, 0}$\cr
$c(S_m, S_n)=\alpha\cdot \frac1m\delta_{m+n, 0}$\cr
\hline 
$c(L_m, S_n)=(\gamma+\beta\cdot m)\delta_{m+n, 0}$\cr
$c(E^i_m, F^i_n)=(\frac12\gamma+\beta\cdot (m+ \frac12))\delta_{m+n+1, 0}$\cr
\hline 
$c(T^{ij}_m, T^{kl}_n)=-\beta\cdot m\delta_{m+n, 0}, \; \text{where}\; (i, j, k, l)\in
A_4$.\cr
\hline 
\end{tabular}
$$
To express the cocycle in terms of {\it primary} fields (i.e., the elements of
$\fwitt$-modules such that the cocycle does not vanish on the pair of elements from one
module only), let us embed
$\fwitt$ differently and, simultaneously, suitably intermix the odd generators:
$$
\begin{matrix}
\tilde L_m=L_m+a_mS_m \; \text{for}\; a_m=-\frac\beta\alpha\cdot
m^2-\frac\gamma\alpha\cdot m;\\
\tilde E^i_m=E^i_m+b_mF^i_m \; \text{for}\;
b_m=\frac{\beta}{2\alpha}\cdot(2m+1)+\frac{\gamma}{2\alpha}.\\
\end{matrix}
$$

In the new basis the cocycle is of the form:
$$
\renewcommand{\arraystretch}{1.4}
\begin{tabular}{|c|}
\hline
$c(\tilde L_m, \tilde L_n)=\left(\frac{\alpha^2-\beta^2}{\alpha}\cdot
m^3-\frac{\alpha^2-\gamma^2}{\alpha}\cdot
m\right)\delta_{m+n, 0}$\cr 
$c(\tilde E^i_m, \tilde E^i_n)=\left(\frac{\alpha^2-3\beta^2}{\alpha}\cdot
(m+\frac12)^2-\frac{\alpha^2-3\gamma^2}{4\alpha}\right)\delta_{m+n+1,
0}$\cr 
$c(T^{ij}_m, T^{ij}_n)=\alpha\cdot m\delta_{m+n, 0}$\cr
$c(F^i_m, F^i_n)=\alpha\cdot \delta_{m+n+1, 0}$\cr
$c(S_m, S_n)=\alpha\cdot \frac1m\delta_{m+n, 0}$\cr
\hline 
$c(\tilde E^i_m, F^i_n)=(\gamma+\beta\cdot (2m+1))\delta_{m+n+1, 0}$\cr
\hline 
$c(T^{ij}_m, T^{kl}_n)=-\beta\cdot m\delta_{m+n, 0}, \; \text{where}\; (i, j, k, l)\in
A_4$.\cr
\hline 
\end{tabular}
$$
It depends on the three parameters and is expressed in terms of primary fields. Observe that
the 3-dimensional space of parameters is not $\Cee^3=\{(\alpha, \beta, \gamma)\}$ but
$\Cee^3$ without a line, since $\alpha$ can never vanish. 

\section*{\S 4. The explicit relations between the Chevalley generators of
\protect $\fk\fas^L$}

Let $\Lambda ^k$ be the
subsuperspace of $\fk^L(1|6)$ generated by the $k$-th degree monomials in the odd
indeterminates $\theta_i$. Then the basis elements of $\fk^L(1|6)$ with their degrees wrt to
$K_t$ is given by the following table:
$$
\renewcommand{\arraystretch}{1.2}
\begin{tabular}{|c|c|c|c|c|c|c|}
\hline
...&$-2$&$-1$&0&1&2&...\cr
\hline 
...&1                    &       &$t$                    &         &$t^2$&...\cr
...&                     &$\Lambda$&                     &$t\Lambda$&        &...\cr
...&$\frac{\Lambda^2}{t}$&       &$\Lambda^2$&        &$t\Lambda^2$&...\cr
...&                     &$\frac{\Lambda^3}{t}$&                     &$\Lambda^3$&       
&...\cr ...&$\frac{\Lambda^4}{t^2}$&       &$\frac{\Lambda^4}{t}$&        &$\Lambda^4$&...\cr
...&                     &$\frac{\Lambda^5}{t^2}$& &$\frac{\Lambda^5}{t}$&  &...\cr 
...&$\frac{\Lambda^6}{t^3}$&   &$\frac{\Lambda^6}{t^2}$&     &$\frac{\Lambda^6}{t}$&...\cr
\hline 
\end{tabular}
$$

\ssec{4.1. The Chevalley generators in $\fk\fas^L$ in terms of $\fo(6)$}
Explicitly, in terms of the generating functions, the basis elements of $\fk\fas^L$ are given
by the following formulas, where $\Theta=\xi_1\xi_2\xi_3\eta_3\eta_2\eta_1$,
$\hat\eta_i=\xi_i$ and
$\hat\xi_i=\eta_i$. Let $\tilde T^{ij}$ ($i=1, dots , 6$) be the matrix skew-symmetric with
respect to the side diagonal with only $(i,j)$-th and $(j,i)$th nonzero entries equal to
$\pm 1$; let
$\tilde G^i=\theta_i$, where
$\theta=(\xi_1, \xi_2, \xi_3, \eta_3, \eta_2, \eta_1)$. Let the
$\tilde S$ denote the generators of one of the two irreducible components in the 
$\fo(6)$-module
$\Lambda^3(\id)$. We will later identify $\tilde G$ with the space
of skew-symmetric $4\times 4$ matrices and $\tilde S$ with the
$\fsl(4)$-module
$S^2(\id)$ of symmetric $4\times 4$ matrices, namely,
$\tilde S^{\pm\varepsilon_{i}}$ for $i= 1, 2, 3$ will be the symmetric offdiagonal matrices;
$\tilde S^{2, 0, 0}$, $\tilde S^{-2, 2, 0}$, $\tilde S^{0, -2, 2}$ and $\tilde S^{0, 0, -2}$ the
diagonal matrix units (the superscripts of $\tilde S$ are the weights of the matrix
elements of the symmetric bilinear form wrt
$\fsl(4)$, see sec. 4.2). Set
$$
\renewcommand{\arraystretch}{1.4}
\begin{tabular}{|c|c|}
\hline
the element&its generator\cr
\hline
$L(2n-2)$&$t^{n}-n(n-1)(n-2)t^{n-3}\Theta$,\cr
$\tilde G^i(2n-1)$&$t^{n}\theta_i-n(n-1)t^{n-2}\pderf{\Theta}{\hat\theta_i}$,\cr
$\tilde T^{ij}(2n)$&$t^{n}\theta_i\theta_j-nt^{n-1}\pder{\hat\theta_i}\pder{\hat\theta_j}\Theta$,\cr
$\tilde S^{\varepsilon_{i}}(2n+1)$&$t^{n}\xi_i(\xi_j\eta_j+\xi_k\eta_k)$,\cr
$\tilde S^{-\varepsilon_{i}}(2n+1)$&$t^{n}\eta_i(\xi_j\eta_j-\xi_k\eta_k)$,\cr
$\tilde S^{2, 0, 0}(2n+1)$&$t^{n}\xi_1\xi_2\xi_3$,\cr
$\tilde S^{-2, 2, 0}(2n+1)$&$t^{n}\xi_1\eta_2\eta_3$,\cr
$\tilde S^{0, -2, 2}(2n+1)$&$t^{n}\xi_2\eta_1\eta_3$,\cr
$\tilde S^{0, 0, -2}(2n+1)$&$t^{n}\xi_3\eta_1\eta_2$.\cr
\hline 
\end{tabular} 
$$
where $\tilde G$ and $\tilde S$ are the following skew-symmetric and symmetric matrices,
respectively; we set $\tilde G^i=\tilde G^{\varepsilon_{i}}$, where $\varepsilon_{i}$ and
$-\varepsilon_{i}$ is the weight of $\xi_i$ and $\eta_i$, respectively, wrt $(H^1, H^2,
H^3)\in\fsl(4)$. 
$$
\tilde G=\begin{pmatrix} 
0&-\xi_1&-\xi_2&\eta_3\\
&0&\xi_3&\eta_2\\
&&0&\eta_1\\
&&&0\\
\end{pmatrix},
\quad\tilde S=\begin{pmatrix} 
\xi_1\xi_2\xi_3&\xi_1(\xi_2\eta_2+\xi_3\eta_3)&\xi_2(\xi_1\eta_1+
\xi_3\eta_3)&\eta_3(\xi_1\eta_1-\xi_2\eta_2)\\ 
&\xi_1\eta_2\eta_3&\xi_3(\xi_1\eta_1+\xi_2\eta_2)&\eta_2(\xi_1\eta_1-
\xi_3\eta_3)\\ 
&&\xi_2\eta_1\eta_3&\eta_1(\xi_2\eta_2-\xi_3\eta_3)\\ 
&&&\xi_3\eta_1\eta_2\\ 
\end{pmatrix}.
$$
These generators, expressed via monomial generators of $\fk^L(1|6)$, are rather complicated. Let us
pass to simpler ones using the isomorphism $\fsl(4)\cong\fo(6)$. Explicitly, this
isomorphism is defined as follows:
$$
\begin{pmatrix} 
&\xi_2\eta_3&\xi_1\eta_3&\xi_1\xi_2\\ 
\xi_3\eta_2&&\xi_1\eta_2&-\xi_1\xi_3\\ 
-\xi_3\eta_1&\xi-2\eta_1&&\xi_2\xi_3\\ 
\eta_1\eta_2&\eta_1\eta_3&\eta_2\eta_3&\\ 
\end{pmatrix},\quad \begin{matrix} 
H^1=-(\xi_2\eta_2-\xi_3\eta_3),\\ 
H^2=-(\xi_1\eta_1-\xi_2\eta_2),\\ 
H^3=(\xi_2\eta_2+\xi_3\eta_3).\\ 
\end{matrix}
$$

\ssec{4.2. The multiplication table in $\fk\fas^L$}In terms of $\fsl(4)$-modules we get a
more compact expression of the elements of $\fk\fas^L$. Let
$\ad$ be the adjoint module, $S$ the symmetric square of the identity $4$-dimensional module
$\id$ and $G=\Lambda^2(\id^*)$; let $\Cee\cdot 1$ denote the trivial module. Then the basis
elements of $\fk\fas^L(1|6)$ with their degrees wrt to $K_t$ is given by the following table
in which the degrees of $t$ indicate the grading
$$
\renewcommand{\arraystretch}{1.2}
\begin{tabular}{|c|c|c|c|c|c|c|}
\hline
degree&$-2$&$-1$&0&1&2&...\cr
\hline 
space&$\Cee\cdot t^{-1}$, $\ad\cdot t^{-1}$& $S\cdot t^{-1}$, $G\cdot t^{-1}$&$\Cee\cdot 1$,
$\ad$&$S$,
$G$&$\Cee\cdot t$,
$\ad\cdot t$& ...\cr
\hline 
\end{tabular}
$$
Though it is impossible to embed $\fwitt\supplus\fgl(4)^{(1)}$ into $\fk\fas^L$, it is
convenient to express the brackets in $\fk\fas^L$ in terms of the matrix units of
$\fgl(4)^{(1)}$ that we will denote by
$T^i_j(a)$; we further set $H_1(a)=T^1_1(a)-T^2_2(a)$, $H_2(a)=T^2_2(a)-T^2_2(a)$ and
$H_3(a)=T^2_2(a)-T^3_3(a)$. Clearly, the rhs in the last line of the following multiplication table
can be expressed via the
$H_i(a)$. We denote the basis elements of the trivial $\fsl(4)$-module of degree $a$ by $L(a)$ and
norm them so that they commute as the usual basis elements of $\fwitt$.

The multiplication table in $\fk\fas^L$ is given by the following table:
$$
\begin{tabular}{rcl}
${}[L(a),\;  L(b)]$&=&$(b-a)L(a+b), $\cr  
${}[L(a),\; T^i_j(b)]$&=&$bT^i_j(a+b), $\cr  
${}[L(a),\; S^{ij}(b)]$&=&$(b+\frac 12 a)S^{ij}(a+b), $\cr  
${}[L(a),\; G_{ij}(b)]$&=&$(b-\frac 12 a)G_{ij}(a+b), $\cr  
${}[T^i_j(a),\; T^k_l(b)]$&=&$\delta^k_jT^i_l(a+b)-\delta^i_lT^k_j(a+b), $\cr  
${}[T^i_j(a),\;  S^{kl}(b)]$&=&$\delta^k_jS^{il}(a+b)+\delta^l_jS^{ik}(a+b), $\cr  
${}[T^i_j(a),\;  G_{kl}(b)]$&=&$\delta^i_kG_{lj}(a+b)+\delta^i_lG_{jk}(a+b)+a\sigma(j, k, l,
m)S^{im}(a+b), $\cr  
${}[S^{ij}(a),\;  S^{kl}(b)]$&=&$0, $\cr 
${}[S^{ij}(a),\; G_{kl}(b)]$&=&$
(2\delta^i_kT^j_l-2\delta^i_lT^j_k+2\delta^j_kT^i_l-2\delta^j_lT^i_k)(a+b),$
\end{tabular}
$$
$$
\begin{aligned}
{}[G_{ij}(a),\;
G_{kl}(b)]=2(b-a)\left(\delta_{j,k}\sigma(i, j, l, m)
T^m_j(a+b)+\delta_{i,k}\sigma(i, j, l, m)T^m_i(a+b)+\right.\\
\left.\delta_{j,l}\sigma(i, k, l,
m)T^m_j(a+b)+\delta_{i,l}\sigma(j, i, k, m)T^m_i(a+b)\right)(a+b)+\\
\sigma(i, j, k, l)\left(-4L(a+b)+(b-a)(T^i_i(a+b)+T^j_j(a+b)-T^k_k(a+b)-T^l_l(a+b))\right)
\end{aligned}
$$

\ssec{4.3. The relations between Chevalley generators in $\fk\fas^L$ in terms of $\fsl(4)$}
Denote: $T_{ij}^{a}=T_{j}^i(a)$. For the positive Chevalley generators we take same of
$\fsl(4)=\Span(T_{ij})$ and the lowest weight vectors $S^1_{44}$ and
$G^1_{12}$ of $S^1$ and $G^1$, respectively. For the negative Chevalley generators we take
same of $\fsl(4)$ and the highest weight vectors $S^{-1}_{11}$ and
$G^{-1}_{34}$ of
$S^{-1}$ and
$G^{-1}$, respectively. Then the defining relations, stratified by weight, are the
following ones united with the usual Serre relations in $\fsl(4)$ (we skip them) and the
relations that describe the highest (lowest) weight vectors:

$$
\begin{tabular}{rcl}
${}\left[T_{23}^0,\; \left[T_{23}^0,\; G_{12}^1\right]\right] $&=&$ 0$\cr   
\end{tabular}
\qquad
\begin{tabular}{rcl}
${}\left[T_{34}^0,\; \left[T_{34}^0,\; \left[T_{34}^0,\; S_{44}^1\right]\right]
     \right] $&=&$ 0$\cr 
\end{tabular}
$$

$$
\begin{tabular}{rcl}
${}\left[\left[G_{12}^1,\; \left[T_{23}^0,\; G_{12}^1\right], \right],\; 
      \left[T_{34}^0,\; S_{44}^1\right]\right] $&=&$ 0$\cr
${}\left[S_{44}^1,\; \left[T_{34}^0, \; S_{44}^1\right]\right] $&=&$ 0$\cr
\end{tabular}
$$
$$
\begin{tabular}{rcl}
${}\left[\left[G_{12}^1,\; \left[T_{23}^0,\; T_{34}^0\right]
      \right],\; \left[\left[T_{23}^0,\; G_{12}^1\right],\; 
      \left[T_{34}^0,\; S_{44}^1\right]\right]\right] $&=&$ 0$\cr
${}\left[\left[G_{34}^{-1},\; \left[T_{23}^0,\; T_{34}^0\right]
      \right],\; \left[\left[T_{23}^0,\; G_{34}^{-1}\right],\; 
      \left[T_{34}^0,\; S_{44}^1\right]\right]\right] $&=&$ 0$\cr
\end{tabular}
$$
$$
\begin{tabular}{rcl}
${}[L^0,\; S_{11}^{-1}] $&=&$ - S_{11}^{-1}$\cr  
$ {}[L^0,\; G_{34}^{-1}] $&=&$ - G_{34}^{-1}$\cr  
${}[H_1^0,\; S_{11}^{-1}] $&=&$ 2  S_{11}^{-1}$\cr  
${}[H_1^0,\; G_{34}^{-1}] $&=&$ 0$\cr  
${}[H_2^0,\; S_{11}^{-1}] $&=&$ 0$\cr  
${}[H_2^0,\; G_{34}^{-1}] $&=&$ G_{34}^{-1}$\cr  
${}[H_3^0,\; S_{11}^{-1}] $&=&$ 0$\cr  
${}[H_3^0,\; G_{34}^{-1}] $&=&$ 0$
\end{tabular}
\qquad
\begin{tabular}{rcl}
${}[L^0,\; S_{44}^1] $&=&$ S_{44}^1$\cr  
${}[L^0,\; G_{12}^1] $&=&$ G_{12}^1$\cr  
${}[H_1^0,\; S_{44}^1] $&=&$ 0$\cr  
${}[H_1^0,\; G_{12}^1] $&=&$ 0$\cr  
${}[H_2^0,\; S_{44}^1] $&=&$ 0$\cr  
${}[H_2^0,\; G_{12}^1] $&=&$ - G_{12}^1$\cr  
${}[H_3^0,\; S_{44}^1] $&=&$ -2  S_{44}^1$\cr  
${}[H_3^0,\; G_{12}^1] $&=&$ 0$
\end{tabular}
$$
$$
\begin{tabular}{rcl}
${}[T_{12}^0,\; S_{11}^{-1}] $&=&$ 0$\cr  
${}[T_{12}^0,\; G_{34}^{-1}] $&=&$ 0$\cr  
${}[T_{23}^0,\; S_{11}^{-1}] $&=&$ 0$\cr  
${}[T_{23}^0,\; G_{34}^{-1}] $&=&$ 0$\cr  
${}[T_{34}^0,\; S_{11}^{-1}] $&=&$ 0$\cr  
${}[T_{34}^0,\; G_{34}^{-1}] $&=&$ 0$
\end{tabular}
\qquad
\begin{tabular}{rcl}
${}[S_{44}^1,\; T_{21}^0] $&=&$ 0$\cr  
${}[S_{44}^1,\; T_{32}^0] $&=&$ 0$\cr  
${}[S_{44}^1,\; T_{43}^0] $&=&$ 0$\cr  
${}[G_{12}^1,\; T_{21}^0] $&=&$ 0$\cr  
${}[G_{12}^1,\; T_{32}^0] $&=&$ 0$\cr  
${}[G_{12}^1,\; T_{43}^0] $&=&$ 0$
\end{tabular}
$$

$$
\begin{tabular}{rcl}
${}[T_{12}^0,\; S_{44}^1] $&=&$ 0$\cr  
${}[T_{12}^0,\; G_{12}^1] $&=&$ 0$\cr  
${}[T_{23}^0,\; S_{44}^1] $&=&$ 0$\cr  
${}[T_{34}^0,\; G_{12}^1] $&=&$ 0$\cr  
\end{tabular}
\qquad
\begin{tabular}{rcl}
${}[T_{21}^0,\; G_{34}^{-1}] $&=&$ 0$\cr  
${}[T_{32}^0,\; S_{11}^{-1}] $&=&$ 0$\cr  
${}[T_{43}^0,\; S_{11}^{-1}] $&=&$ 0$\cr  
${}[T_{43}^0,\; G_{34}^{-1}] $&=&$ 0$
\end{tabular}
$$

$$
\begin{tabular}{rcl}
${}[S_{44}^1,\; S_{44}^1] $&=&$ 0$\cr  
${}[S_{44}^1,\; G_{12}^1] $&=&$ 0$\cr  
${}[G_{12}^1,\; G_{12}^1] $&=&$ 0$\cr  
${}[S_{11}^{-1},\; S_{11}^{-1}] $&=&$ 0$\cr  
${}[S_{11}^{-1},\; G_{34}^{-1}] $&=&$ 0$\cr  
${}[G_{34}^{-1},\; G_{34}^{-1}] $&=&$ 0$
\end{tabular}
\qquad
\begin{tabular}{rcl}
${}[S_{44}^1,\; S_{11}^{-1}] $&=&$ 0$\cr  
${}[S_{44}^1,\; G_{34}^{-1}] $&=&$ -4  T_{43}^0$\cr  
${}[G_{12}^1,\; S_{11}^{-1}] $&=&$ 4 
T_{12}^0$\cr  $  {}[G_{12}^1,\; G_{34}^{-1}] $&=&$-4 L^0 - 2H_1^0 - 4H_2^0 - 2  H_3^0$
\end{tabular}
$$

\end{document}